\documentclass[twocolumn,times]{aastex631}%

\usepackage[]{newtxtext}
\usepackage[]{newtxmath}

\usepackage{boogaard-commands}

\shorttitle{NOEMA HDF-N Survey}
\shortauthors{Boogaard et al.}

\begin{document}

\title{A NOEMA molecular line scan of the Hubble Deep Field North:\\ Improved constraints on the CO luminosity functions and cosmic density of molecular gas}

\correspondingauthor{Leindert Boogaard}
\email{boogaard@mpia.de}

\author[0000-0002-3952-8588]{Leindert A. Boogaard} \affiliation{Max
  Planck Institute for Astronomy, K\"onigstuhl 17, 69117 Heidelberg,
  Germany}

\author[0000-0002-2662-8803]{Roberto Decarli}
\affiliation{INAF-Osservatorio di Astrofisica e Scienza dello Spazio,
  via Gobetti 93/3, I-40129, Bologna, Italy}

\author[0000-0003-4793-7880]{Fabian Walter} \affiliation{Max Planck
  Institute for Astronomy, K\"onigstuhl 17, 69117 Heidelberg, Germany}
\affiliation{National Radio Astronomy Observatory, Pete V. Domenici
  Array Science Center, P.O. Box O, Socorro, NM 87801, USA}

\author[0000-0003-4678-3939]{Axel Wei\ss} \affil{Max-Planck-Institut
  f\"{u}r Radioastronomie, Auf dem H\"{u}gel 69, 53121 Bonn, Germany}

\author[0000-0003-1151-4659]{Gerg\"{o} Popping} \affil{European Southern
  Observatory, Karl-Schwarzschild-Str. 2, D-85748, Garching, Germany}

\author[0000-0002-7176-4046]{Roberto Neri} \affil{Institut de
  Radioastronomie Millim\'{e}trique (IRAM), 300 Rue de la Piscine, 38400
  Saint-Martin-d’H\`{e}res, France}

\author[0000-0002-6290-3198]{Manuel Aravena} \affil{N\'ucleo de Astronom\'ia de
  la Facultad de Ingenier\'ia y Ciencias, Universidad Diego Portales,
  Av. Ej\'ercito Libertador 441, Santiago, Chile}

\author[0000-0001-9585-1462]{Dominik Riechers}
\affil{I. Physikalisches Institut, Universit\"{a}t zu K\"{o}ln,
  Z\"{u}lpicher Strasse 77, 50937, K\"{o}ln, Germany}

\author[0000-0001-7782-7071]{Richard S. Ellis}\affil{Department of
  Physics and Astronomy, University College London, Gower Street,
  London WC1E 6BT, UK}

\author[0000-0001-6647-3861]{Chris Carilli} \affil{National Radio Astronomy Observatory, Pete
  V. Domenici Array Science Center, P.O. Box O, Socorro, NM 87801, USA}

\author[0000-0003-2027-8221]{Pierre Cox} \affil{Institut d'astrophysique de
  Paris, Sorbonne Université, CNRS, UMR 7095, 98 bis bd Arago, 7014 Paris,
  France}

\author[0000-0003-3061-6546]{J\'{e}r\^{o}me Pety} \affil{Institut de
  Radioastronomie Millim\'{e}trique (IRAM), 300 Rue de la Piscine, 38400
  Saint-Martin-d’H\`{e}res, France}

\begin{abstract}
  We present measurements of the CO luminosity functions (LFs) and the
  evolution of the cosmic molecular gas density out to $z\sim6$ based
  on an 8.5\,arcmin$^{2}$ spectral scan survey at 3\,mm of the iconic
  Hubble Deep Field North (HDF-N) observed with the NOrthern Extended
  Millimeter Array (NOEMA).  We use matched filtering to search for
  line emission from galaxies and determine their redshift probability
  distributions exploiting the extensive multi-wavelength data for the
  HDF-N.  We identify the 7 highest-fidelity sources as CO emitters at
  $1<z<6$, including the well-known submillimeter galaxy HDF\,850.1 at
  $z=5.18$.  Four high-fidelity 3\,mm continuum sources are all found
  to be radio galaxies at $z\leq1$, plus HDF\,850.1.  We constrain the
  CO LFs in the HDF-N out to $z\sim6$, including a first measurement
  of the CO(5--4) LF at $\avg{z}=5.0$.  The relatively large area
  and depth of the NOEMA HDF-N survey extends the existing luminosity
  functions at $1<z<4$ above the knee, yielding a somewhat lower
  density by 0.15--0.4\,dex at the overlap region for the CO(2--1) and
  CO(3--2) transitions, attributed to cosmic variance.  We perform a
  joint analysis of the CO LFs in the HDF-N and Hubble Ultra Deep
  Field from ASPECS, finding that they can be well described by a
  single Schechter function.  The evolution of the cosmic molecular
  gas density from a joint analysis is in good agreement with earlier
  determinations.  This implies that the impact of cosmic
  field-to-field variance on the measurements is consistent with
  previous estimates, adding to the challenges for simulations that
  model galaxies from first principles.
\end{abstract}


\section{Introduction} \label{sec:intro}

As star formation takes place inside clouds of cold molecular gas
\citep{McKee2007}, the cosmic density of molecular gas (\rhomol) plays a key
role in our understanding of what drives the cosmic star formation
rate density \citep{Madau2014} and the baryon cycle of matter flowing
in and out of galaxies \citep{Walter2020,Peroux2020}.

Measurements of the cosmic molecular gas density have matured over the
last decade, in particular through so-called spectral scan surveys in
the (sub-)millimeter regime of extragalactic deep fields with large
interferometers.  By scanning for emission from the low-$J$
transitions of carbon monoxide (CO)---one of the key tracers of cold
molecular gas in the local universe \citep{Solomon1992,
  Bolatto2013}---a flux-limited census of the molecular gas reservoirs
in galaxies can be obtained, provided the physical conditions of the
systems under study are known.

The first constraints on the cosmic molecular gas density from
individual CO detections were obtained over a 1\,arcmin$^2$ area in
the Hubble Deep Field North \citep{Williams1996} in the 3\,mm band
with the Plateau de Bure Interferometer
\citep[PdBI,][]{Walter2014,Decarli2014}.  Building on these, the ALMA
Spectroscopic Survey of the Hubble Ultra Deep Field (ASPECS)-Pilot
program targeted a $\sim 1$\,arcmin$^2$ area in the Hubble Ultra Deep
Field \citep{Beckwith2006} in both the 3\,mm and 1.2\,mm bands at
greater depth
\citep{Walter2016,Aravena2016b,Aravena2016c,Bouwens2016a,Decarli2016a,Decarli2016b,Carilli2016}.
These initial efforts led to the first extragalactic ALMA large
program, ASPECS, that covered the entire eXtreme Deep Field region of
the HUDF \citep{Illingworth2013} over a 4.6\,arcmin$^2$ scan at 3\,mm
\citep{Aravena2019,Boogaard2019,Boogaard2021,
  Decarli2019,Gonzalez-Lopez2019,Popping2019,Uzgil2019} and 1.2\,mm
\citep{Aravena2020,Boogaard2020,Bouwens2020,Decarli2020,
  Gonzalez-Lopez2020,Inami2020,Popping2020,Magnelli2020}.  At the same
time a significantly larger $\sim 60$\,arcmin$^2$ area in the COSMOS
and GOODS-North fields was targeted as part of the COLDz survey at
9\,mm \citep{Pavesi2018,Riechers2019} on the Karl G. Jansky Very Large
Array, probing the bright end of the luminosity function for lower-$J$
CO transitions at similar or higher redshifts.

Through the larger area and complementary line and redshift
coverage, these surveys have provided an increasingly detailed picture
of the cosmic molecular gas density out to $z\sim4$.  These reveal
that $\rhomol(z)$ increases by a factor $\sim 6\times$ going out to
$z\sim1.5$, with a subsequent decline out to the highest redshifts
probed \citep{Walter2020}.  However, the speed at which spectral scan
surveys can be conducted is limited by the bandwidth that can be
observed per tuning ($\sim 4$\,GHz per sideband for ALMA) and the
requirement to perform mosaicing over larger areas.  This implies that
the deepest surveys to date are still probing volumes that are subject
to cosmic variance \citep{Decarli2020,Popping2020}, though the
three-dimensional volumes probed are significantly larger than the
modest on-sky areas may suggest.

To address these issues, we have conducted a new survey using the
NOrthern Extended Millimeter Array (NOEMA), covering the full Hubble
Deep Field North in a 45 pointing mosaic encompassing 8.5\,arcmin$^2$
at 3\,mm.  Capitalising on the quadrupled instantaneous bandwidth of
the POLYFIX correlator (compared to the original PdBI scan) of 16\,GHz
and the increased sensitivity from four extra antennas in the array,
we cover almost the full 3\,mm window between 82---113\,GHz in only 2
setups.  For comparison, the NOEMA HDF-N survey covers almost
$\sim 2\times$ the area of ASPECS at a $\sim 3\times$ shallower depth.

This paper is organised as follows. In \autoref{sec:observations} we
present the observations and data reduction.  We discuss the line
search, the identification of both the line and continuum emitters,
and the subsequent computation of the luminosity functions (LF) in
\autoref{sec:results}, also including a comparison to the original
PdBI scan.  We discuss constraints on the CO LFs and the cosmic
molecular gas density in \autoref{sec:discussion}.  We summarize and
conclude in \autoref{sec:conclusion}.  Throughout this paper, we adopt
a \cite{PlanckCollaboration2018a} cosmology (flat $\Lambda$CDM with
$H_{0} = 67.66$\,km\,s$^{-1}$\,Mpc$^{-1}$, $\Omega_{m} = 0.3111$ and
$\Omega_{\Lambda} = 0.6889$).  We use $\log$ to denote $\log_{10}$ and
$\ln$ for the natural logarithm.

\begin{deluxetable}{cccccr}[t]
  \tablecaption{Lines and corresponding redshift ranges and volume covered in the HDF-N mosaic.
    \label{tab:freqzvol}}
  \tablehead{
    \colhead{Transition} & \colhead{$\nu_{\rm rest}$} & \colhead{$z_{\rm min}$} & \colhead{$z_{\rm max}$} & \colhead{$\langle z \rangle$} & \colhead{Volume} \\
    \colhead{} & \colhead{($\mathrm{GHz}$)} & \colhead{} & \colhead{}  & \colhead{} & \colhead{($\mathrm{Mpc^{3}}$)} \\
  \colhead{(1)} & \colhead{(2)} &\colhead{(3)} & \colhead{(4)} &\colhead{(5)} &\colhead{(6)} }
  \startdata
  CO(1--0)   & 115.271 & 0.0173 & 0.399 & 0.2936 & 995 \\
  CO(2--1)   & 230.538 & 1.0345 & 1.798 & 1.4389 & 19963 \\
  CO(3--2)   & 345.796 & 2.0517 & 3.1969 & 2.6276 & 34591 \\
  CO(4--3)   & 461.041 & 3.0687 & 4.5956 & 3.8210 & 42446 \\
  CO(5--4)   & 576.268 & 4.0856 & 5.9941 & 5.0156 & 46693 \\
  CO(6--5)   & 691.473 & 5.1023 & 7.3923 & 6.2106 & 49016 \\
  CO(7--6)   & 806.652 & 6.1188 & 8.7902 & 7.4056 & 50251 \\
  \hline
  \CI(1--0)  & 492.161 & 3.3434 & 4.9733 & 4.1436 & 43857 \\
  \CI(2--1)  & 809.342 & 6.1425 & 8.8228 & 7.4335 & 50271 \\
  \enddata
  \tablecomments{The frequency coverage ranges from 82.394 --
    113.322\,GHz.  The comoving volume and volume-weighted average
    redshifts are computed within 0.5 of the primary beam peak
    sensitivity (8.5 arcmin$^2$ at 98 GHz), accounting for its
    frequency dependence.}
\end{deluxetable}
\begin{figure}[t]
\includegraphics[width=\columnwidth]{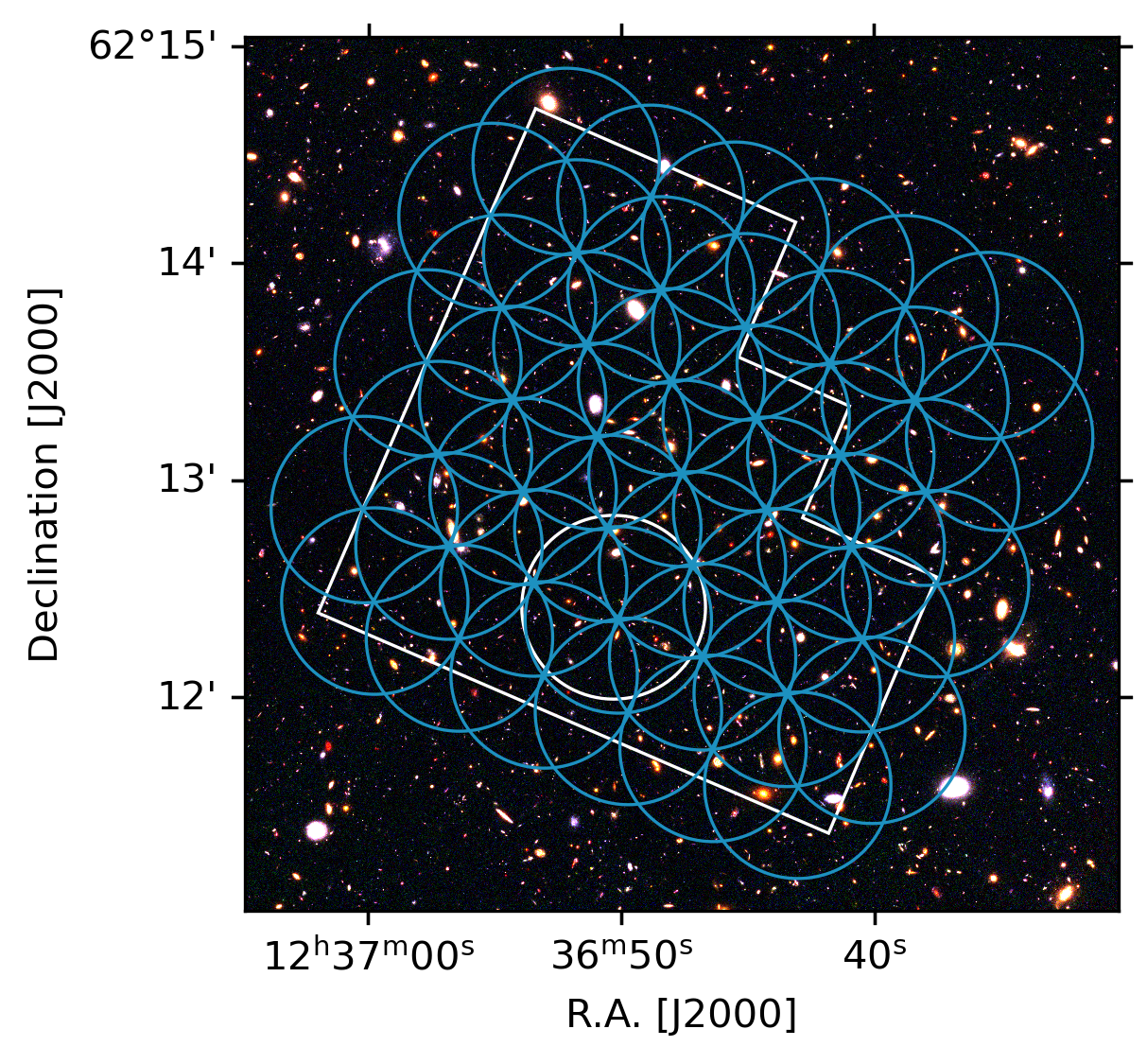}
\caption{Footprint of the 45-pointing NOEMA mosaic around the central
  frequency of 98\,GHz (in blue) compared to the footprint of the
  original Hubble Deep Field North observations with WFPC2 from
  \citet[][in white]{Williams1996}.  The white circle indicates the
  original Plateau de Bure scan \citep{Walter2014, Decarli2014}.
  The background shows the Hubble imaging from CANDELS
  \citep{Grogin2011, Koekemoer2011} in the WFC3/F160W, ACS/F850LP, and
  ACS/F606W filters (RGB).\label{fig:footprint}}
\end{figure}
\begin{deluxetable}{ccccc}
  \tablecaption{Properties of the dirty cubes and continuum map. \label{tab:cubeprops}}
  \tablehead{
    \colhead{Tuning} & \colhead{$\nu_0$} & \colhead{Beam size}            & \colhead{Beam P.A.} & \colhead{$\langle \mathrm{rms} \rangle$} \\
    \colhead{}       & \colhead{(GHz)}   & \colhead{(arcsec$^{2}$)}                     & \colhead{(degrees)}  & \colhead{(mJy\,beam$^{-1}$)} \\
  \colhead{(1)} & \colhead{(2)} &\colhead{(3)} & \colhead{(4)} &\colhead{(5)} }
\startdata
Setup 1 LSB          & 86.300            & \(4.82 \times 4.33\) & 87.9               & 0.42 \\
Setup 2 LSB          & 94.026            & \(4.44 \times 3.92\) & 90.2                & 0.60 \\
Setup 1 USB          & 101.600           & \(4.11 \times 3.68\) & 86.5                & 0.53 \\
Setup 2 USB          & 109.362           & \(3.75 \times 3.33\) & 91.0                & 0.95 \\
Continuum            & 97.822            & \(4.26 \times 3.75\) & 87.4                & 0.011 \\
\enddata
\tablecomments{(1) The four sidebands, plus the continuum. (2) Central
  frequency. (3) Beam size. (4) Beam position angle. (5)
  Root-mean-square (rms) noise, which for the cubes is the average rms
  per 9\,MHz channel.}
\end{deluxetable}

\section{NOEMA Observations}
\label{sec:observations}
The NOEMA mosaic consists of 45 pointings that are laid-out in a
Nyquist-sampled hexagonal pattern at around 98\,GHz, with the phase
center set to 12:36:47.60 +62:13:02.0.  It covers 8.5\,arcmin$^{2}$
(50\% peak sensitivity at 98\,GHz) in the GOODS-North area,
encompassing the complete Hubble Deep Field North
(\autoref{fig:footprint}). The mosaic was observed in two setups that
cover nearly the full 3\,mm band from 82.394--113.322\,GHz.  The main
emission lines that are covered in spectral scan and their associated
redshift range and cosmic volume are listed in \autoref{tab:freqzvol}.

The observations were taken between the 27th of March and 20th April
2019 for setup 1 and between 9th of May and 15th of October 2020 for
setup 2.  The calibration of the mosaic was performed in
\textsc{clic}.  For setup 1, a total of 7 tracks were used in the
final reduction. The calibrators were 3C273 and 3C84 for the bandpass,
1125+569 and J1302+690 for the amplitude and phase (using the average
polarisation for the amplitude when detected), and LKHA101 and MWC349
for the absolute flux calibration, except for the track on the 2nd of
April 2019, for which 1055+018 and 1125+569 were used as the bandpass
and flux calibrators respectively.  For setup 2, a total of 10 tracks
were used, with bandpass calibrators 3C273, 0851+202, 3C84, and 3C345,
amplitude and phase calibrator 1125+569, and flux calibrators MWC349
and 3C84.%

We create a dirty cube of the entire mosaic for each of the four
sidebands separately, with 9\,MHz channels and a $0\farcs75$ pixel
size, using \textsc{Gildas} (version August 22a).  The reference
frequency is set to the center of each sideband and we take into
account frequency dependence of the synthesised beam for every channel
by explicitly setting $\texttt{map\_beam\_step} = 1$ in
\texttt{uvmap}.  We subsequently compute the (frequency-dependent)
sensitivity map of the mosaic (or `primary beam of the mosaic') as the
weighted primary beam response of individual pointings.  We use the
sensitivity map to create both (flux-calibrated) cubes where the noise
is flat (that are used for the line search) and
sensitivity-map-corrected cubes (with noise increasing towards the
edges of the mosaic).  The beam shape and root-mean-square (rms) noise
of each of the four cubes is detailed in \autoref{tab:cubeprops}.

\begin{figure}[t]
\includegraphics[width=\columnwidth]{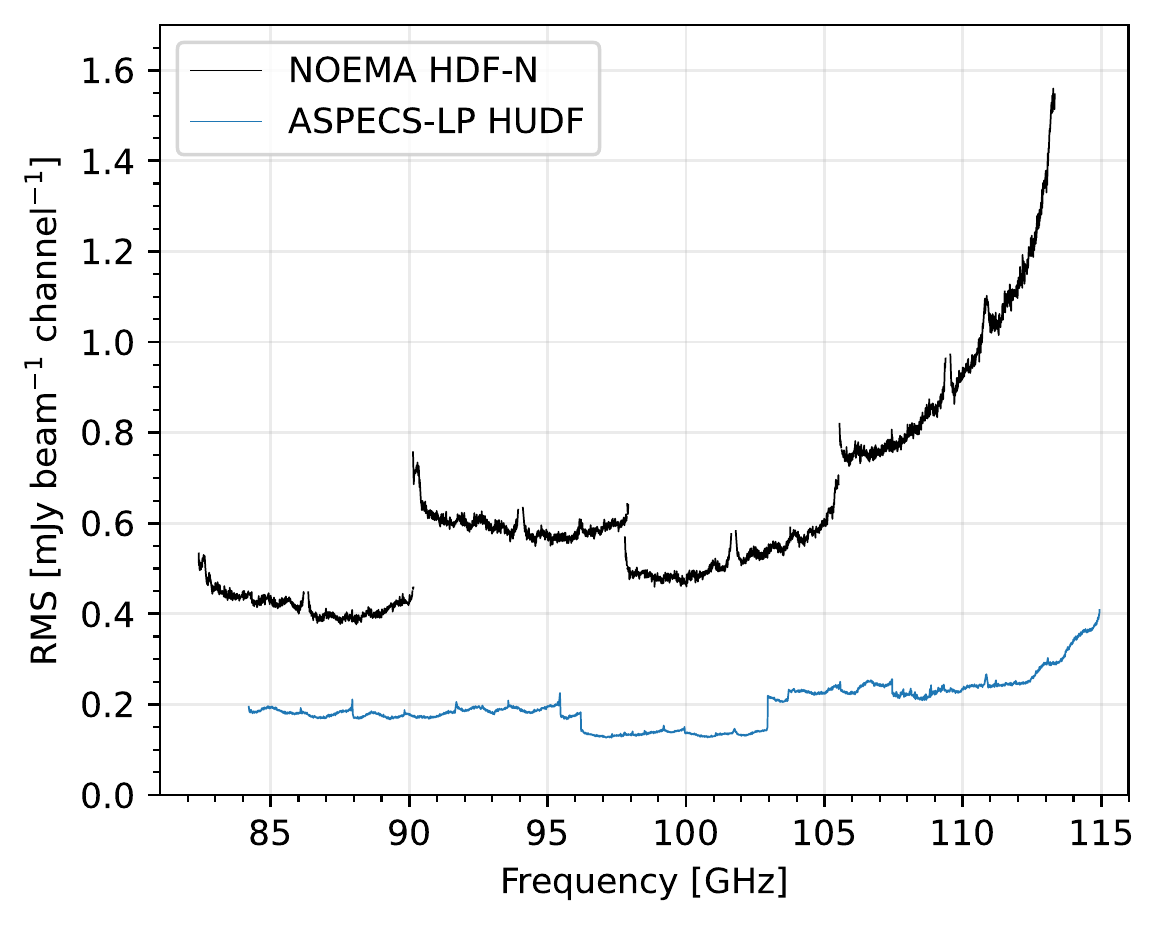}
\caption{The root-mean-square noise (rms) as a function of frequency.
  The rms is measured per 9\,MHz channel and the upper and lower
  sidebands of each of the two setups of the NOEMA HDF-N scan are
  shown separately.  For comparison, the rms of the ASPECS-LP
  \citep[][measured per 7.8 MHz channel]{Decarli2019} is shown in
  blue.\label{fig:rms}}
\end{figure}
\begin{deluxetable*}{cccccccccccc}
  \tabletypesize{\footnotesize}
  \tablecaption{Highest fidelity lines from the line search and their
    identification.  \label{tab:linesearch}} \tablehead{
    \colhead{ID} & \colhead{RA}      & \colhead{Decl.}   & \colhead{S/N} & \colhead{Fidelity} & \colhead{PB} & \colhead{Frequency}           & \colhead{FWHM} & \colhead{Integrated flux}             & \colhead{$J$} & \colhead{$z$} & \colhead{comment}\\
    \colhead{}   & \colhead{(J2000)} & \colhead{(J2000)} & \colhead{}    & \colhead{}         & \colhead{}   &    \colhead{($\mathrm{GHz}$)} & \colhead{(km\,s$^{-1}$)} & \colhead{(Jy\,km\,s$^{-1}$)} & \colhead{}    & \colhead{}    & \colhead{} \\
    \colhead{(1)} & \colhead{(2)} &\colhead{(3)} & \colhead{(4)}
    &\colhead{(5)} & \colhead{(6)} & \colhead{(7)} &\colhead{(8)} &
    \colhead{(9)} &\colhead{(10)} &\colhead{(11)} & \colhead{(12)} }
  \startdata
  1 & 12:36:34.52 & +62:12:41.0 & 12.38 & 1.00 & 0.48 & 103.655 $\pm$ 0.008 & 518 $\pm$ 58  & 2.14 $\pm$ 0.21 & 2           & 1.225      & spec-$z$ \\
  2 & 12:36:48.57 & +62:12:16.2 & 7.72  & 1.00 & 0.99 & 82.777 $\pm$ 0.006  & 325 $\pm$ 53  & 0.50 $\pm$ 0.07 & 2           & 1.785      & D14.ID3, spec-$z$  \\
  3 & 12:36:52.00 & +62:12:26.0 & 6.89  & 1.00 & 0.99 & 93.191 $\pm$ 0.011  & 417 $\pm$ 83  & 0.58 $\pm$ 0.10 & 5           & 5.184      & HDF\,850.1, spec-$z$ \\
  4 & 12:36:40.73 & +62:14:06.5 & 6.40  & 0.99 & 0.69 & 104.060 $\pm$ 0.014 & 447 $\pm$ 94  & 0.76 $\pm$ 0.14 & 3$^{\dagger}$       & 2.323$^{\dagger}$ &  $p(z)$  ambiguous\\
  5 & 12:36:33.01 & +62:13:41.0 & 6.01  & 0.83 & 0.52 & 83.518 $\pm$ 0.016  & 567 $\pm$ 135 & 0.88 $\pm$ 0.18 & 3$^{\dagger}$       & 3.141$^{\dagger}$ & $p(z)$  ambiguous \\
  6 & 12:36:44.81 & +62:12:07.2 & 5.99  & 0.73 & 1.00 & 109.173 $\pm$ 0.003 & 74  $\pm$ 17  & 0.29 $\pm$ 0.06 & 2$^{\dagger}$           & 1.112$^{\dagger}$      & $p(z)$ peak ($>90\%$) \\
  7 & 12:36:38.80 & +62:12:57.5 & 5.85  & 0.81 & 0.98 & 107.494 $\pm$ 0.007 & 171 $\pm$ 44  & 0.40 $\pm$ 0.09 & 2           & 1.145      & spec-$z$ \\
  \enddata \tablecomments{(1) Line ID.  (2--3) Right ascension and
    declination.  (4) S/N from matched filtering. (5) Fidelity. (6)
    Mosaic sensitivity relative to peak at the source position. (7-9)
    Line frequency, full-width at half-maximum (FWHM) and integrated
    flux from a Gaussian fit. (10) Upper $J$ level of identified CO
    transition. (11) Redshift. (12) Comment on identification.
    $^{\dagger}$$p(z)$ peak redshift solution, cf.\ \autoref{fig:pz}.}
\end{deluxetable*}

We mask individual channels with increased noise, such as those at the
edges of the two basebands in each sideband, by identifying all
channels that deviate by more than 5\% from the median filtered rms
over 50 channels (in total $<4\%$ of all channels).  The resulting rms
in each sideband is shown in \autoref{fig:rms} and the overall average
rms over the full frequency coverage is 0.62
mJy\,beam$^{-1}$\,channel$^{-1}$.  Because of the differences in the
rms and beam sizes between the individual sidebands, we do not create
a single combined cube.  Instead, throughout the remainder of this
paper, we perform all analysis (i.e., line searches, completeness
corrections) on each of the four individual cubes separately.

We also create a map of the 3\,mm continuum map by combining all
setups, after masking channels with increased noise in the same way as
for the cubes, albeit more aggressively (using a 5\% cut with a
200\,channel median filter).  The rms noise is
11\,$\mu$Jy\,beam$^{-1}$ and the beam shape is listed alongside other
properties in \autoref{tab:cubeprops}.
\begin{figure*}[t]
  \includegraphics[width=\textwidth]{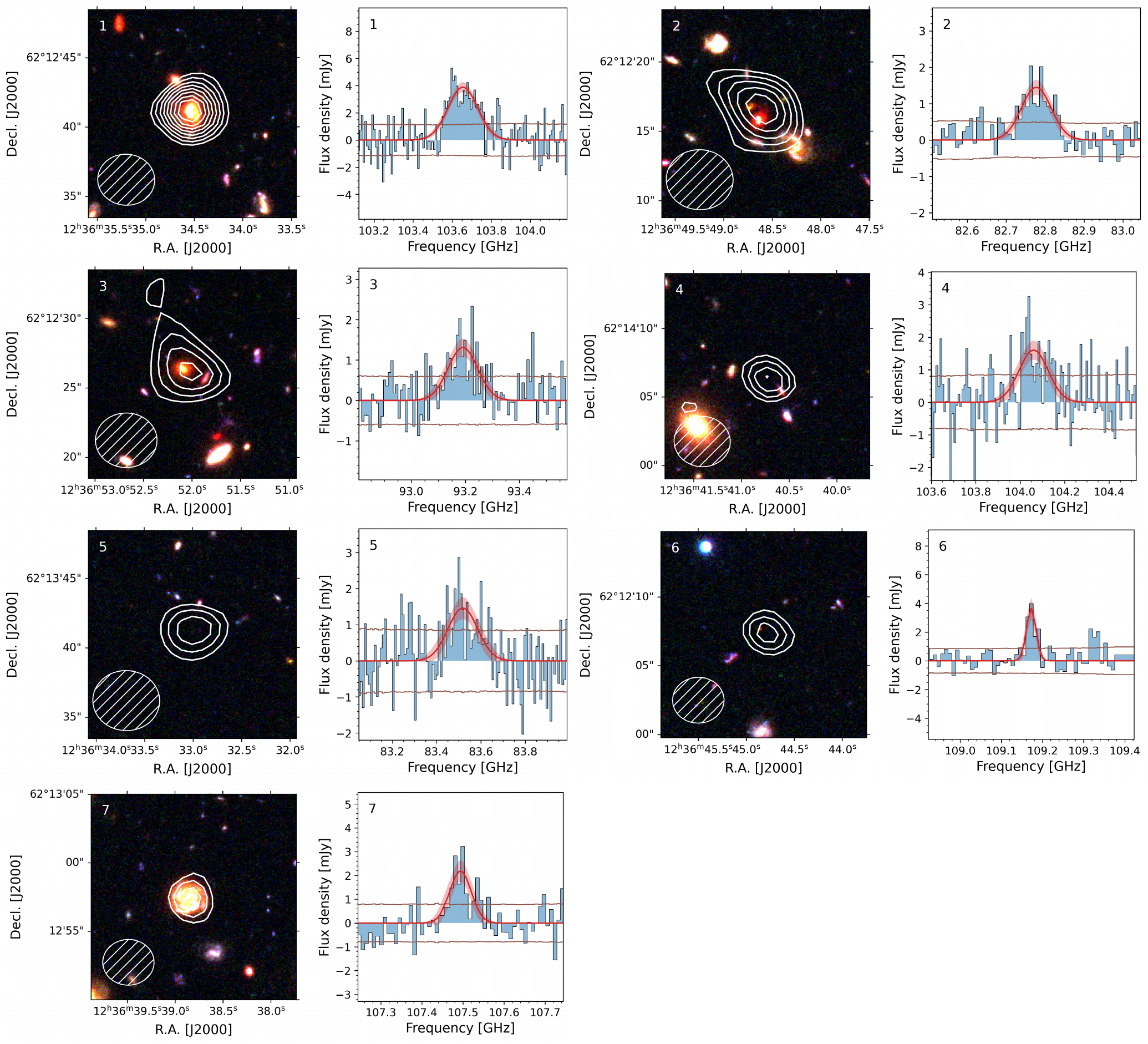}
  \caption{Emission line candidates with highest fidelity and \SN.
    The left panels show the moment zero maps overlaid on the HST
    images (F160W/F850LP/F606W).  Contours are spaced from
    $\pm3\sigma$ up to $\pm10\sigma$, spaced by $\pm1\sigma$ (no
    negative contours are seen) and the synthesised beam is shown in
    the bottom left corner.  The right panels show the peak pixel
    spectra and the rms noise (brown line), including Gaussian fits
    and $1\sigma$-uncertainties (red line and shading).}
\end{figure*}

\section{Analysis and Results} \label{sec:results}
\subsection{Line Search}
\label{sec:linesearch}

We search the cubes for positive and negative line emission using
matched filtering.  Several codes have been developed in the context
of different spectral scan surveys, including \textsc{Findclumps}
\citep{Walter2016, Decarli2019}\footnote{Now implemented in
  \textsc{Interferopy} \citep{interferopy};
  \url{https://github.com/interferopy/interferopy}.},
\textsc{Lineseeker}\footnote{\url{https://github.com/jigonzal/LineSeeker}}
\cite{Gonzalez-Lopez2019} and
\textsc{MF3D}\footnote{\url{https://github.com/pavesiriccardo/MF3D}}
\citep{Pavesi2018}.  All these codes have been found to perform
qualitatively similarly, though differ somewhat in the way they
perform the matched filtering and group the final list of candidates
\citep[e.g.,][]{Gonzalez-Lopez2019}.

As our fiducial line search code, we use \textsc{MF3D}, which conducts
the matched filtering on the \SN\ cube using spatial and spectral
kernels of different sizes.  We use a Gaussian kernel in frequency
space, with widths ranging from 3 to 18 channels, corresponding to
line widths of about 75 to 500\,km\,s$^{-1}$ (FWHM).  We use a
single-pixel (point-like) spatial kernel, as all sources are expected
to be unresolved at the beam size.

We use the distribution of the negative lines (expected to be produced
by noise), to estimate a `fidelity' of the positive line signal as a
function of \SN.  The fidelity is computed per kernel ($\sigma$) by
taking the ratio of the number of positive and negative lines
($N_{\rm pos}$, $N_{\rm neg}$) in bins of \SN,
\begin{align}
  F(\SN, \sigma) = 1 - \frac{N_{\rm neg}(\SN, \sigma)}{N_{\rm pos}(\SN, \sigma)}\label{eq:fid}
\end{align}
To mitigate the effect of low-number statistics on the estimate of
$N_{\rm neg}(\SN, \sigma)$, we fit the counts with a tail of a
Gaussian function centered at zero (as in \citealt{Gonzalez-Lopez2019,
  Decarli2019, Decarli2020}).  We finally compute a smooth estimate of
the $F(\SN)$ at fixed $\sigma$ by fitting an error function shape to
\autoref{eq:fid} \citep[cf.][]{Walter2016}.  As the number of sources
detected in the individual cubes is rather limited for a given kernel
width, the estimate of the fidelity at a fixed \SN\ is uncertain
(except at the tails of the distribution, close to zero and unity
fidelity). To mitigate this effect, we combine the line search results
in S/N-space and compute a single fidelity estimate for a given kernel
width for all cubes combined.  This is equivalent to performing the
analysis on the combined cube, as is typically done.  This provides a
more robust estimate of the fidelity, though we still caution against
over-interpreting the exact fidelity values in the intermediate range.

For the final catalog we only consider lines candidates with a
$\SN > 4$ (also when remeasured with a Gaussian fit, see
\autoref{sec:properties}), a $F>0.2$ (consistent with earlier work),
and that lie in the area of the mosaic where the sensitivity is above
40\% of the peak sensitivity.  This leaves a total of 23 sources, all
of which have a $\SN \ge 5$ due to the fidelity cut being more
stringent than the \SN\ cut, and that lie within $\sim 50\%$ of
primary peak sensitivity.  There is a relatively sharp drop in
fidelity below $\sim 0.7$ and the majority of the sources lie at lower
fidelity- and \SN-end.  We list the highest fidelity sources, with
$F \geq 0.7$, in \autoref{tab:linesearch}, which have a
$\SN \geq 5.85$.

We estimate the completeness of our line search by injecting simulated
emission lines into the cubes and determining what fraction is
recovered by our line search procedure.  We assume a 3D Gaussian
profile for the simulated lines, which matches shape of the beam for
each cube in the spatial directions.  We draw 1000 sources from a
uniform distribution in peak flux and FWHM line width, ranging from 0
to 3\,mJy and 0 to 800\,km\,s$^{-1}$ respectively.  The sources are
then injected uniformly across the cubes in the area above 50\% of the
peak sensitivity.  We perform the line search in the same manner as
described above and define the completeness ($C$) as the fraction of
the injected sources that are recovered for a given line width and
peak flux.  We repeat the experiment 5 times for each cube to increase
statistics to a total of 5000 lines per sideband (20.000 simulated
lines in total).  The resulting completeness fractions are shown in
\autoref{fig:compl} in \autoref{sec:colf-impact}.  In the end, we find
the completeness corrections are minor, as the uncertainties are
dominated by the sample purity (fidelity) and not completeness.

We cross check the line candidates found with \textsc{MF3D} against
those found by \textsc{Findclumps} and \textsc{Lineseeker}.  Overall,
we recover the same high-\SN\ and -fidelity candidates with all the
codes (though there are differences in the absolute \SN, due to the
different methods), while there are increasing numbers of candidates
found by only a subset of the codes at lower \SN.  Similar conclusions
were also reached by \citet{Decarli2019} and
\citet{Gonzalez-Lopez2019}.  We further assess the impact of the
line-search code used in \autoref{sec:colf-impact}, where we show we
recover the same CO luminosity functions if we use the line candidates
from \textsc{Findclumps} in favor of MF3D.

\subsection{Counterpart association}
\label{sec:count-assoc}
\begin{figure}[t]
  \centering
  \includegraphics[width=0.8\columnwidth]{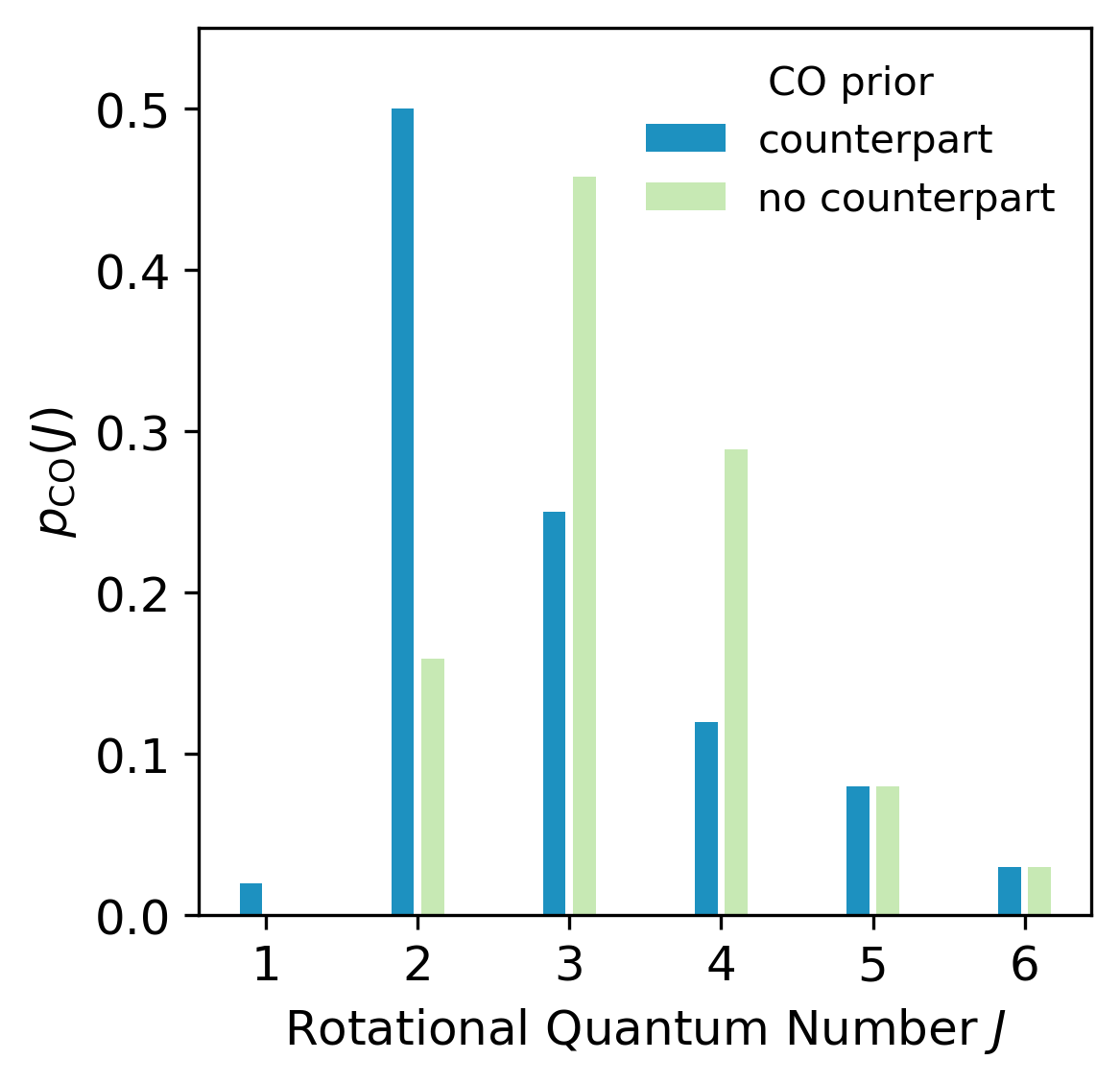}
  \caption{
    CO line prior used for the redshift
    associations.  The prior for sources with an (HST) counterpart is
    based on the observed CO line distribution from ASPECS
    \citep{Decarli2020}.  For the sources without counterpart, we use
    a line distribution based on the redshift distribution of
    optically-faint submillimeter galaxies from \cite{Smail2021} in
    combination with a typical CO ladder \citep{Danielson2011,
      Boogaard2020}.
    \label{fig:priors}}
\end{figure}

We identify the redshifts of the candidates from the line-search by
exploiting the extensive multi-wavelength data that is available over
the HDF-N as compiled by the CANDELS and 3D-HST surveys
\citep{Grogin2011, Koekemoer2011, Brammer2012, Skelton2014,
  Momcheva2016}, including spectroscopic redshifts
\citep[e.g.][]{Barger2008}, as well as the \emph{Herschel} data
\citep{Elbaz2011}.

While for the brightest and/or lower-redshift galaxies the counterpart
association is often clear, the relatively large beam of the NOEMA
data (compared to typical galaxy sizes) can sometimes lead to
ambiguities in the association. There may be multiple galaxies near
the peak of the emission or, especially at lower S/N, the peak of the
emission may be offset from the position of the counterpart.
Moreover, the counterpart may not be detected in the
optical/near-infrared imaging at all.

To deal with these uncertainties, we assign a redshift probability for
every line candidate, based on the photometric redshift distributions
of galaxies in the vicinity, weighted by their relative distance to
the source.  We also include a `dark' solution, in which there is no
optical/NIR photometric counterpart.  We use the photometric redshift
distributions that are derived with \textsc{eazy} by 3D-HST
\citep{Brammer2012, Skelton2014, Momcheva2016}.  We take a prior on
the separation between the line candidate and the photometric
counterpart in the form of a Gaussian with a 2'' FWHM (about half the
beam size), i.e., a radial down-weighting of sources at a larger
separation from the line position.

\begin{figure}[t]
\includegraphics[width=\columnwidth]{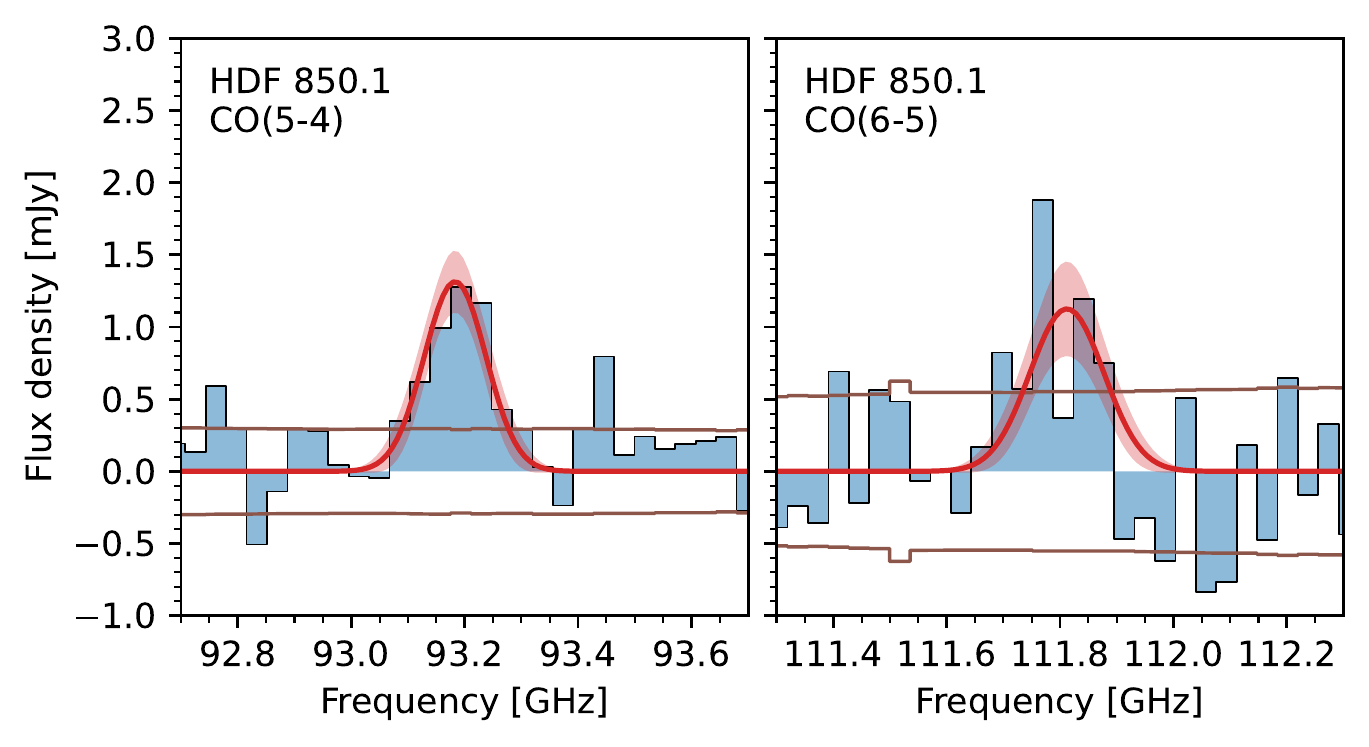}
\caption{Spectra of CO(5-4) and CO(6-5) in HDF\,850.1, binned by a
  factor 5.  Both lines are detected, but CO(6-5) only at lower S/N
  (below the line-search threshold).  A joint Gaussian fit to both
  lines is shown in red, yielding integrated fluxes of
  $S_{\mathrm{CO(5-4)}} = 0.58 \pm 0.10$\,Jy\,km\,s$^{-1}$,
  $S_{\mathrm{CO(6-5)}} = 0.49 \pm 0.14$\,Jy\,km\,s$^{-1}$, and a line
  width of $412\pm72$\,km\,s$^{-1}$ at
  $z=5.184\pm0.001$.\label{fig:hdf850}}
\end{figure}

Because it is well known that not all CO transitions are observed in
equal numbers at fixed observing frequency, we take into account
different prior probabilities on the CO line identification.  For the
sources with a photometric counterpart, we adopt a prior that is
loosely based on the line-flux distribution from ASPECS at 3\,mm
\citep[][]{Decarli2020}.  For the sources without a photometric
counterpart, we adopt the redshift distribution of optically-faint
submillimeter galaxies ($K$-faint) as found by \cite{Smail2021} and
convert it to a CO line distribution in $J$ assuming a fiducial CO
ladder based on the typical integrated line flux ratios in SMGs
(\citealt{Danielson2011}, cf.\ \citealt{Boogaard2020}).  Both CO line
priors are shown in \autoref{fig:priors}.  Above a redshift of
$\sim 3$ there is some ambiguity in the line identification, where
multiple lines from CO and \CI\ are potentially present in the
spectrum.  For the identification, we assume that the line is always
the strongest line visible at the respective redshift, and check for
fainter lines afterwards.  This concerns \CI(1--0), that is typically
weaker than CO(4--3) \citep[e.g.,][]{Valentino2020a}, and the $J>6$ CO
lines, that are rarely significantly stronger than CO(6--5)
\citep{Carilli2013}.  Effectively, this means that all lines are
assumed to be CO with $J\leq6$.  We discuss the impact of these priors
on the final results in more detail in \autoref{sec:colf-impact}.

The redshift probability for a line candidate observed at a frequency
$\nu$ and position $s$ is given by:
\begin{align}
  p(z | \nu, s) \propto \sum_i \left[p_i(z | s_i) p(z | \nu) p(s_{i} | s)\right] + p_d(z | \nu, s_d) p(s_d| s), \label{eq:pz}
\end{align}
where the sum is taken over all galaxies that lie within a 6''
diameter circle (i.e., significantly larger than the beam and typical
galaxy sizes).  Here $p_i(z | s_i)$ is the photometric redshift
distribution for galaxy $i$ at position $s_{i}$, $p(z|\nu)$ is the
redshift prior determined by the CO line prior and the observed
frequency, and $p(s_{i}|s)$ is the radial weighting that depends only
on the absolute separation $|s_{i} - s|$.  The last term represent the
no-counterpart or `dark' case, where $p_d(z | \nu, s_{d})$ is the
associated (prior) redshift probability distribution and
$p(s_{d} | s)$ accounts for the relative weight that is given to the
no-counterpart solution.  For the latter we assume the value of the
radial weighting at $2''$, which means the no-counterpart solution
gets a larger weight than solutions with a counterpart for separations
larger than $2''$.  The overall probability is normalised to unity for
each line candidate, depending on the number of photometric
counterparts and their radial weight.

The $p(z)$ distribution for the top sources are shown in
\autoref{fig:pz} in \autoref{sec:redsh-distr}.  We cross-check the
solutions for the highest fidelity sources with the additional
spectroscopic redshifts from literature and find that the $p(z)$
solutions are in perfect agreement with the known spectroscopic
redshifts analysis in all cases.  The redshifts and associated line
identifications are reported in \autoref{tab:linesearch}.

We find the majority of lines (4/7) are CO(2--1) emitters with
redshifts $1 < z < 2$.  Three out of 4 are supported by a
spectroscopic redshift, while for ID.6 the $p(z)$ solution contains
more than 90\% of the total probability.  The high number of CO(2--1)
emitters is very consistent with the findings from ASPECS, where the
majority of lines were from CO(2--1) (roughly 60\%), followed by
CO(3--2) (roughly 30\%), cf.\ \autoref{fig:priors}.  There is no
CO(3--2) emitter with an unambiguous redshift identification, though
the redshift probabilities of both sources with a broader $p(z)$ peak
at $J=3$.  We detect CO(5--4) in HDF\,850.1 and its spectrum also
reveals reveals CO(6--5), albeit at lower fidelity and \SN\ (below the
line-search threshold), shown in \autoref{fig:hdf850}.

An alternative method to identify the CO line redshifts is to use the
long-wavelength dust spectral energy distribution and compare the
inferred dust temperature of the line emitter to the dust temperature
distribution of sources with known redshifts
\citep[e.g.,][]{Weiss2013, Strandet2016}.  To this end, we cross-match
our line list to the Herschel catalog from \cite{Elbaz2011}.  We find
a clear Herschel counterpart within $1''$ for all sources, with the
exception of IDs 4, 5 and 6, hence no additional constraints on their
potential redshift can be derived in this way.

\subsection{Comparison to earlier PdBI observations}
\label{sec:comppdbi}

We re-examine the 16/21 candidate emission lines from the earlier PdBI
observations in the HDF-N \citep[][with a similar average rms of
$\sim 0.55$\,mJy per 9\,MHz channel]{Walter2014, Decarli2014} that
fall within the present frequency coverage (i.e., excluding their
ID.1, 2, 19, 20 and 21).  We confirm their ID.3 and ID.8 (HDF\,850.1),
which are recovered as ID.2 and ID.3 in the present line search.  In
addition, we also confirm ID.17, which is recovered at low fidelity in
the present line search, but is robust, corresponding to CO(6--5) in
HDF\,850.1 (see \autoref{fig:hdf850}).  For the remaining original
line candidates no significant emission is identified in the new
observations.

\subsection{Continuum}
\label{sec:continuum}
\begin{figure}[t]
\includegraphics[width=\columnwidth]{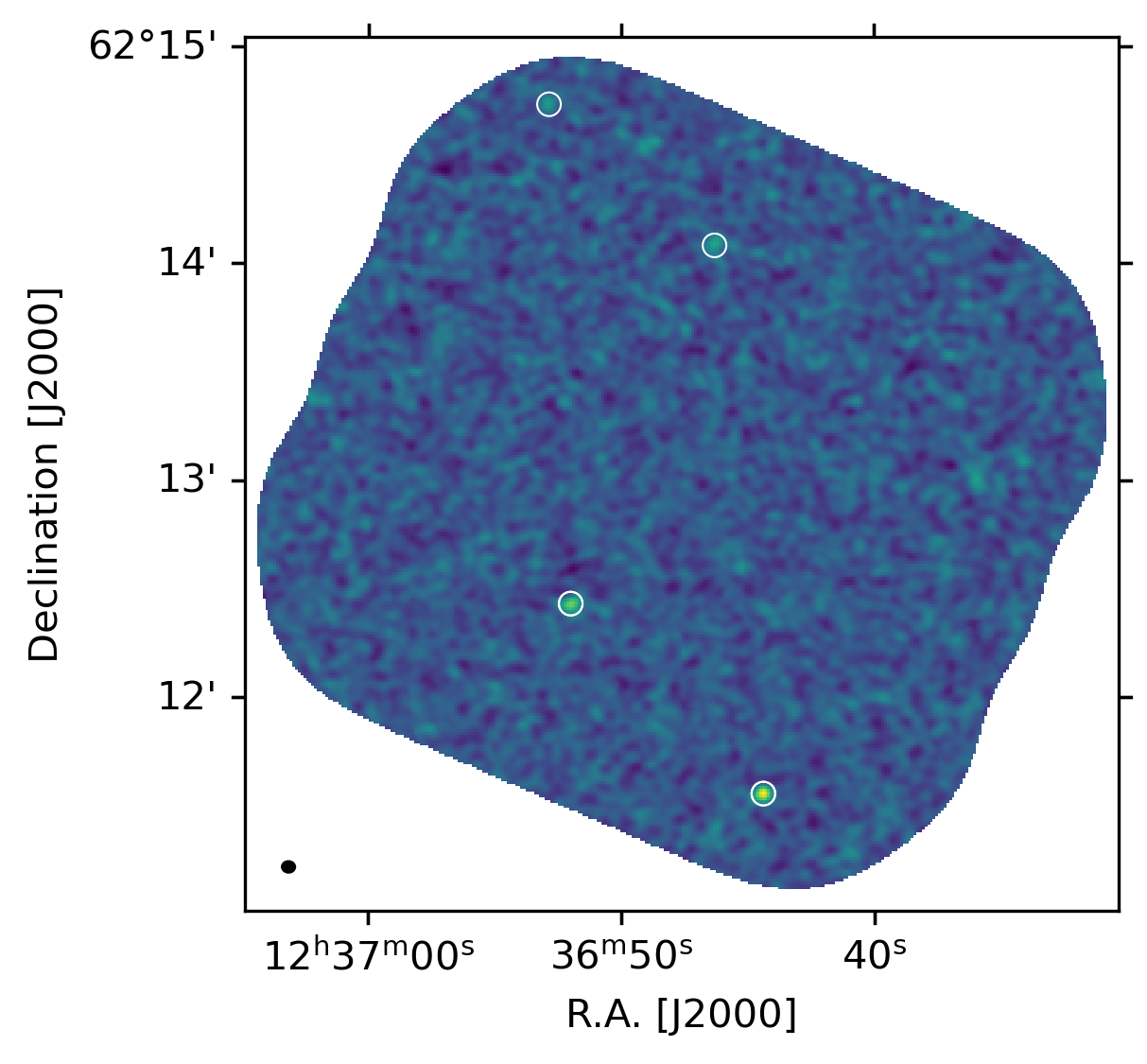}
\caption{Continuum map at 97.8\,GHz before applying the mosaic primary
  beam correction.  The rms noise is 11\,$\mu$Jy\,beam$^{-1}$ and the
  beam size is shown in the bottom left corner.  The brightest
  sources are marked with circles (cf.\
  \autoref{fig:cont_src}).\label{fig:continuum}}
\end{figure}
\begin{deluxetable}{cccccccc}
  \tabletypesize{\footnotesize}
  \tablecaption{Properties of continuum sources. \label{tab:cont}}
  \tablehead{\colhead{ID} & \colhead{R.A.} & \colhead{Decl.} & \colhead{S/N} & \colhead{F} & \colhead{PB} & \colhead{$S_{\nu, 3mm}$} & \colhead{$z$}\\
    \colhead{ } & \colhead{(J2000) } & \colhead{(J2000)} & \colhead{ }  & \colhead{ } & \colhead{ } & \colhead{($\mathrm{\mu Jy}$)} \\
\colhead{(1)} & \colhead{(2)} &\colhead{(3)} & \colhead{(4)} &\colhead{(5)} & \colhead{(6)} & \colhead{(7)} &\colhead{(8)}
}  \startdata
  C1 & 12:36:44.38 & +62:11:33.5 & 11.01 & 1.00 & 0.78 & $148 \pm 22$ & 1.013 \\
  C2 & 12:36:52.00 & +62:12:26.0 & 7.55 & 1.00 & 0.99  & $110 \pm 23$ & 5.184 \\
  C3 & 12:36:46.31 & +62:14:05.0 & 4.74 & 0.92 & 0.93  & $ 92 \pm 17$ & 0.961 \\
  C4 & 12:36:52.86 & +62:14:44.0 & 4.32 & 0.81 & 0.52 & $ 103 \pm 25$ &
  0.321
  \enddata
  \tablecomments{(1) Continuum ID.  (2--3) Right ascension and
    declination.  (4) S/N from matched filtering (5) Fidelity. (6)
    Mosaic sensitivity relative to peak at the source position.  (7)
    Flux density at 3\,mm. (8) Spectroscopic redshifts from
    \cite{Barger2008}, except for C2 \citep[HDF\,850.1; ][this
    work]{Walter2012}.}
\end{deluxetable}
\begin{figure*}[t]
\includegraphics[width=\textwidth]{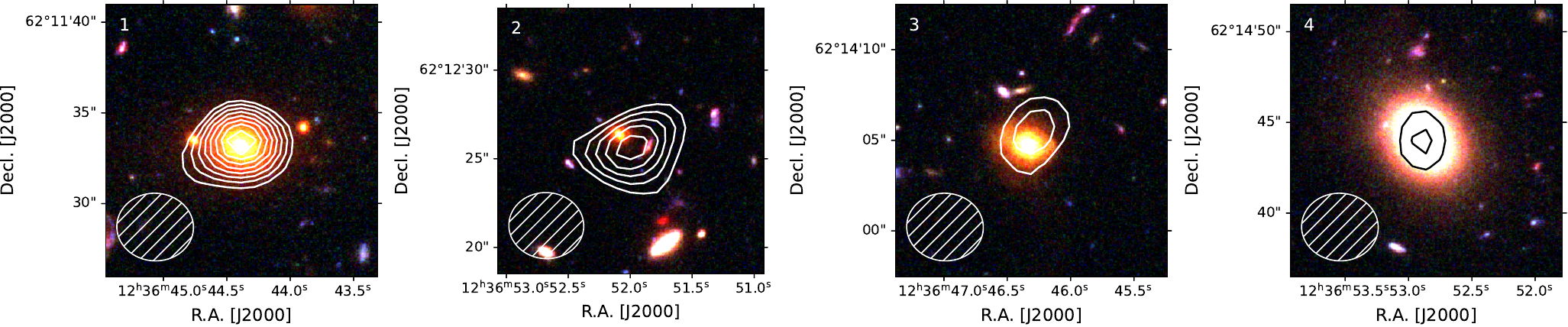}
\caption{Continuum sources overlaid on the HST imaging
  (F160W/F850LP/F606W).  Contours are spaced from $\pm3\sigma$ up to
  $\pm10\sigma$, spaced by $\pm1\sigma$ and the synthesised beam is
  shown in bottom left corner.  The second continuum source is
  HDF\,850.1, the other three are all radio galaxies at $z\leq1$ (see
  references in text).\label{fig:cont_src}}
\end{figure*}
The 3\,mm continuum map is shown in \autoref{fig:continuum}, prior to
the mosaic primary beam-correction (i.e., with flat noise).  We search
for sources in the dirty continuum map using matched filtering in the
same way as for the cubes, but with a 1-channel spectral template.  We
detect four continuum sources with a fidelity above 0.8 ($\SN > 4.3$).
We Hogbom clean the cube around the brightest sources from the line
search down to $2\sigma$.  We measure the fluxes in the cleaned image
by fitting 2D Gaussians using \texttt{imfit} in \textsc{casa}.  The
properties of the sources and measured fluxes are listed in
\autoref{tab:cont}.

We show HST cutouts of the continuum sources in
\autoref{fig:cont_src}.  The second brightest source corresponds to
HDF\,850.1.  The other three sources are all identified as known
(radio) galaxies detected at 1.4, 5, 10 and 34\,GHz
\citep{Morrison2010, Owen2018, Gim2019, Murphy2017, Algera2021}
showing AGN signatures in their X-ray \citep{Xue2016} and/or radio
emission \citep{Algera2021}, with known spectroscopic redshifts
$\leq 1$ \citep{Barger2008}.  No bright lines fall within the
frequency range of the mosaic at the redshift of these three sources
(cf.\ \autoref{tab:freqzvol}).  The 3\,mm number counts at $5\sigma$
of $N(>0.09\,\mathrm{mJy}) \approx 850\,\mathrm{deg}^{-2}$ (no
completeness or flux-boosting correction correction) are in good
agreement with the recent estimates from \cite{Zavala2021}.

\subsection{Properties of sources}
\label{sec:properties}

We fit the lines recovered in the line search with a Gaussian line
profile using \textsc{lmfit} \citep{lmfit_1_0_0}.  The resulting line
fluxes, frequencies, and widths are reported for the high-fidelity
sources in \autoref{tab:linesearch}.  We do not apply any flux
boosting corrections, as the completeness simulations
(\autoref{sec:linesearch}) show that this only has a minor impact on
the line flux at the S/N levels under consideration.  As in
\cite{Decarli2020}, we also do not correct the line fluxes for the
impact of the increasing temperature of the Cosmic Microwave
Background (CMB) with redshift, though note these would only
significantly decrease the observed line flux at relatively low
intrinsic excitation temperatures.  We refer the reader to
\cite{Decarli2020} for a more in-depth discussion of both topics.  We
compute the line luminosities ($L'$) in units of
K\,km\,s$^{-1}$\,pc$^{2}$ \citep[e.g.,][]{Solomon1992, Carilli2013}
for each of the the various redshift solutions from the $p(z)$
analysis, which are used for the computation of the CO LFs.

\subsection{Luminosity Function Analysis}
\label{sec:lumin-funct-analys}
\begin{figure*}[t]
\includegraphics[width=\textwidth]{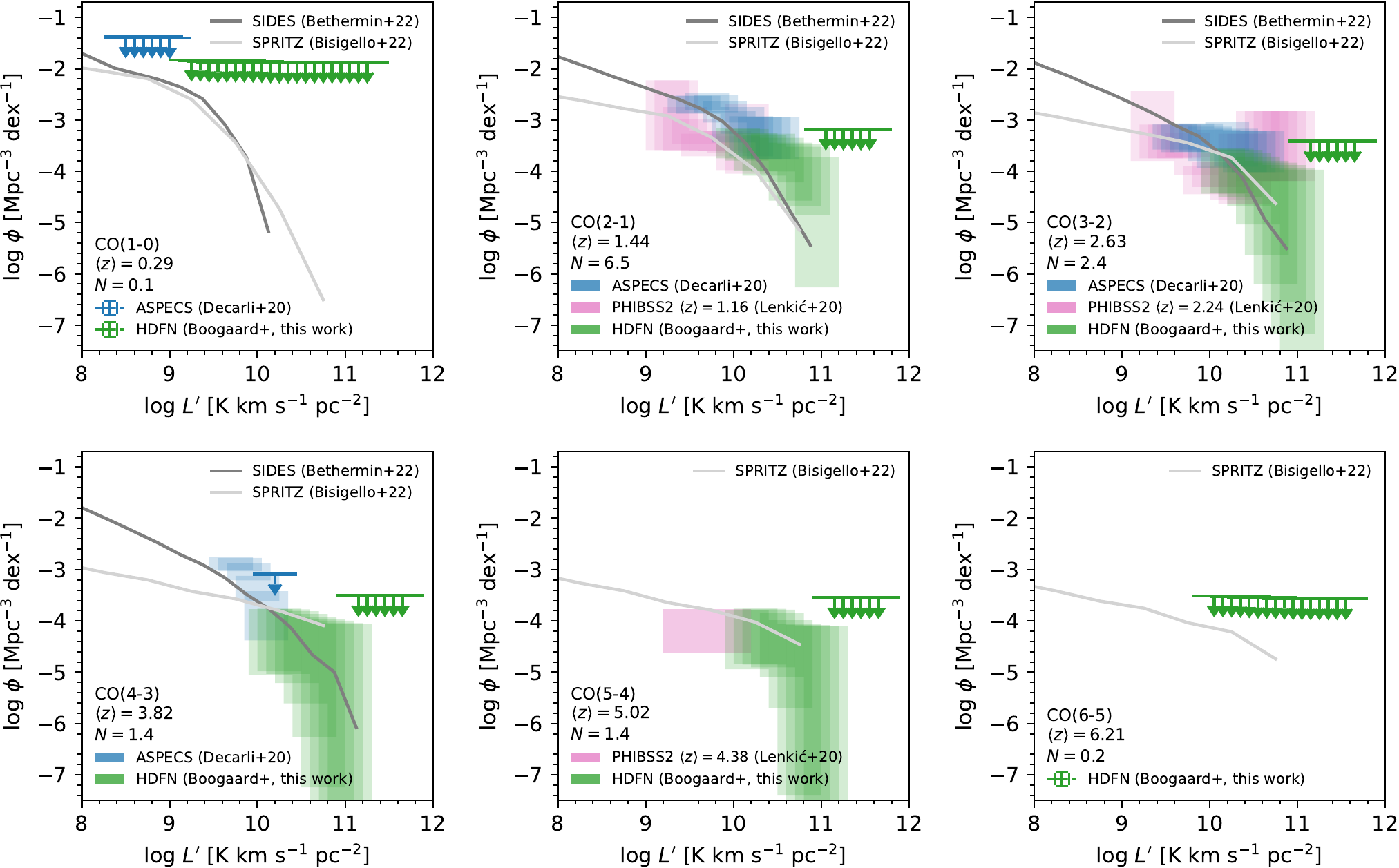}
\caption{CO luminosity functions in the HDFN (green boxes).  Each
  panel indicates the CO transition and mean redshift, as well as the
  mean number of sources that entered the luminosity function (from
  5000 realisations).  The blue boxes show the results from ASPECS
  \citep[][]{Decarli2020} and the pink boxes from PHIBSS2
  \citep{Lenkic2020}.
  \label{fig:colf}}
\end{figure*}
The luminosity functions are computed following the approach of
\cite{Decarli2016a, Decarli2019, Decarli2020}, but taking all 23
sources (\autoref{sec:linesearch}) into account for the luminosity
functions of all the different lines, with a weight depending on their
$p(z)$ solution.  The exception to this are the high-fidelity sources
with a spectroscopic redshift, for which we only include the actual
redshift with a fidelity of unity.

The luminosity function for a certain transition is then defined as:
\begin{align}
  \phi(\log L')~[\mathrm{Mpc}^{-3}\,\mathrm{dex}^{-1}] = \frac{1}{V \Delta(\log L')} \sum_{i=1}^{N} \frac{F_{i}}{C_{i}}. \label{eq:lf}
\end{align}
Here $\phi$ is the number of sources per comoving Mpc$^{3}$ in an
interval of $\log L' \pm 0.5 \log L'$, $V$ is the volume over which a
transition is detectable, $\Delta(\log L')$ is the bin width, and
$F_i$ and $C_i$ are the fidelity and completeness for a given
transition.  To construct the luminosity functions, 5000 independent
realisations are created, where in each realisation the line
luminosities are varied within errors.  The number of sources and
associated $1\sigma$ Poissonian confidence intervals
\citep{Gehrels1986} are computed in 0.5\,dex bins, unless a bin
contains less than one source on average, in which case a $3\sigma$
upper limit is provided.  The resulting counts and uncertainties are
finally scaled by the completeness corrections, divided by the volume
and averaged over the realisations.  Following earlier work, the
luminosity functions are computed five times with offsets of 0.1 dex
(which are therefore not independent), to expose the intra-bin
variations given the modest statistics.

The luminosity functions in the HDFN are shown in \autoref{fig:colf}
for CO(1--0) up to CO(6--5) and are tabulated in \autoref{tab:colfs}.  We
compare these to the recent luminosity functions for the HUDF
determined by the ASPECS Large Program \citep{Decarli2019,
  Decarli2020}.  These are also determined from a spectral scan at
3\,mm, over almost exactly the same redshift interval
(cf.\ \autoref{fig:rms}), but probe down to fainter luminosities over a
$\sim 2.5\times$ smaller volume.  We also show the LFs derived from
background sources in the PHIBSS2 fields \citep{Lenkic2020} for the
redshift ranges that are reasonably close to those from NOEMA HDFN and
ASPECS.

We compare the observed LFs to the recent theoretical predictions from
the SIDES \citep{Bethermin2022} and \textsc{Spritz}
\citep[][]{Bisigello2022} simulations.  In brief, both simulations use
empirical prescriptions for the IR luminosity and CO-to-IR scaling
relations.  SIDES simulates the galaxy population using a
semi-analytical model that that is coupled to the stellar mass of dark
matter halos determined via abundance matching on a dark matter-only
light-cone from the Bolshoi-Planck simulations
\citep{Rodriguez-Puebla2016}, while \textsc{Spritz} is a fully
empirical model that is based on the observed galaxy stellar mass and
IR luminosity functions.

\begin{figure*}[t]
\includegraphics[width=\textwidth]{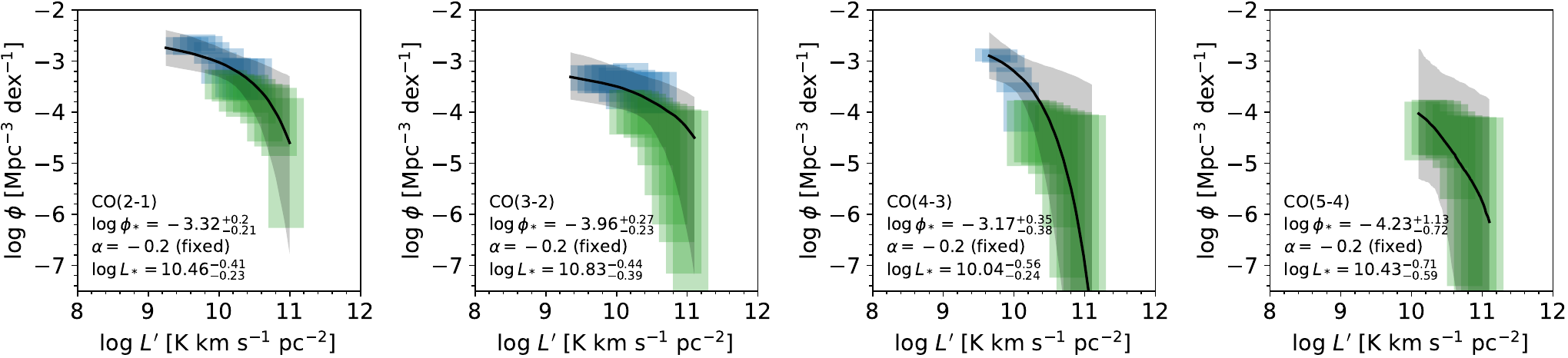}
\caption{Schechter fits to the CO luminosity functions of the HDFN
  (green boxes) + ASPECS (blue boxes).  Each panel indicates the CO
  transition and the fit parameters   (cf.\ \autoref{tab:sfparams}). \label{fig:colf-sf}}
\end{figure*}

In both the local universe and at higher redshift, the (lower-$J$)
luminosity functions have been found to be reasonably well described
by \cite{Schechter1976} functions \citep[e.g.,][]{Saintonge2017,
  Fletcher2020, Riechers2019, Decarli2019}.  Given the broad overall
agreement between the LF's from the HDF-N and HUDF, further discussed
in \autoref{sec:co-lumin-funct}, we perform a joint fit using a
Schechter function in logarithmic units \citep[e.g.,][]{Riechers2019}:
\begin{align}
  \log \phi = \log \phi_{*} + \alpha \log\left(\frac{L'}{L'_{*}}\right) - \frac{1}{\ln 10} \frac{L'}{L'_{*}} + \log(\ln 10) \label{eq:sf}
\end{align}
Here $\phi_{*}$ is the normalisation defining the overall density of
galaxies at the characteristic luminosity $L'_{*}$ (in the same units
as \autoref{eq:lf}) and $\alpha$ is the faint-end slope.  We only fit
uncorrelated bins for each dataset, i.e., one-fifth of bins evenly
spread across the full range probed, to avoid underestimating the
uncertainties (note that we find the same median posterior values if
we would fit all the bins, but with narrower posteriors).  We also
chose not to fit the PHIBSS2 data, as they are not derived
consistently with the redshift intervals of the other surveys.  As we
do not provide new constraints on the faint-end, we fix the slope
$\alpha=-0.2$ following \cite{Decarli2020}, that is consistent with
the local value \citep{Saintonge2017}.  We take uniform priors, where
$-6 \leq \log \phi_{*} \leq -2$ and $8.5 \leq \log L'_{*} \leq 11.5$,
and sample the posterior on the parameters using nested sampling with
\textsc{Ultranest} \citep{Buchner2021}.  The resulting fits are shown
in \autoref{fig:colf-sf} and the marginalised estimates for the
parameters are tabulated in \autoref{tab:sfparams}.

As it appears that CO(4--3) may not be well described by a single
Schechter function, we also explored fitting two Schechter functions
to the $J=2$, 3 and 4 LF's (with broader priors of
$-9 < \log \phi_{*, 1, 2} < -1$ and
$8.0 \leq \log L'_{*, 1, 2} \leq 12.0$, and the requirement that
$\log L_{*,1} < \log L_{*, 2}$).  However, we find that in all cases
the ratio ($K$) of the Bayesian evidence ($Z$) supports the single
Schechter fits, with $K = Z_{\rm double} / Z_{\rm single} \leq 0.5$
\citep[][]{Jeffreys1961}.

\label{sec:co-lumin-funct}
\begin{deluxetable}{cccc}[t]
  \tablecaption{Schechter function parameters posterior percentiles
    for different CO transitions.
    \label{tab:sfparams}}
  \tablehead{
    \colhead{Transition}  & \colhead{$\langle z \rangle$} & \colhead{$\log \phi_{*}$}          &  \colhead{$\log L'_{*}$}     \\
    \colhead{}            & \colhead{}                    & \colhead{(Mpc$^{-3}$\,dex$^{-1})$}  & \colhead{(K\,km\,s$^{-1}$\,pc$^{2}$)} \\
    \colhead{(1)} & \colhead{(2)} &\colhead{(3)} & \colhead{(4)}
  }  \startdata
CO(2--1)              & 1.439      &  $ -3.32_{-0.21}^{+0.20}$  &  $ 10.46_{-0.23}^{-0.41}$  \\
CO(3--2)              & 2.628      &  $ -3.96_{-0.23}^{+0.27}$  &  $ 10.83_{-0.39}^{-0.44}$  \\
CO(4--3)              & 3.821      &  $ -3.17_{-0.38}^{+0.35}$  &  $ 10.04_{-0.24}^{-0.56}$  \\
CO(5--4)              & 5.016      &  $ -4.23_{-0.72}^{+1.13}$  &  $ 10.43_{-0.59}^{-0.71}$  \\
  \enddata
  \tablecomments{See \autoref{eq:sf}. We fix $\alpha=-0.2$.}
\end{deluxetable}

\section{Discussion} \label{sec:discussion}
\subsection{CO luminosity functions}
The large volume of the NOEMA survey provides new constraints on the
bright end of the CO LF, as shown in \autoref{fig:colf}.  Compared to
ASPECS, the roughly $3\times$ shallower observations over $2\times$
the area extend the overall constraints past the knee of the LF, while
reaching comparable constraints near the knee.

We find a lower luminosity density in the overlap regions of the LF in
the HDF-N compared to the HUDF, of roughly 0.15\,dex for CO(2--1) and
0.4\,dex for CO(3-2).  It is not immediately clear where these
differences come from, but they do not appear to be methodological
(\autoref{sec:colf-impact}).  More likely they are due to
field-to-field variance.  It is known that there is an overdensity in the
HUDF at $z\sim1.1$ that may bias the CO(2--1) measurements from ASPECS
high \citep{Boogaard2019}. It is unclear if similar over- or
underdensities affect the comparison of the CO(3--2) LF, though we
cannot confidently identify the same fraction of CO(3--2) emitters as
were found in the HUDF.  Even larger variations are seen when
comparing to the LFs from PHIBSS2.  This suggest that the variations
seen between the fields can be attributed to cosmic variance.
Interestingly both SIDES and \textsc{Spritz} seem to predict a lower
luminosity density for CO(2--1) than is observed in both fields,
making it less clear whether this is due to cosmic variance
\citep{Bethermin2022} or a missing ingredient in the models.  Taken
together, the combined measurements from the HDF-N and HUDF are
(still) reasonably well described by a single Schechter function
(\autoref{fig:colf-sf}).  The joint fits provide improved constraints
on the overall shape of the LFs, which now explicitly take into
account the measured cosmic variance between the fields.

For CO(4--3), it appears we find a larger number of sources at the
bright end than may be expected from an extrapolation from ASPECS.
One should note the total number of sources entering the LF here is
very limited: there are only 1 and 2 independent LF bins for ASPECS
and the HDF-N respectively, with very few sources, hence the
differences could simply be due to noise and low-number statistics.
Given there are no CO lines with spectroscopic confirmation entering
this bin, it is also more sensitive to the assumed priors, and only
upper limits can be derived if one limits to the top-sources (see
\autoref{sec:colf-impact} for more details).  We do find that the
bright end of the CO(4--3) LF is consistent with the predictions from
both SIDES and \textsc{Spritz}, while the models predict a lower
luminosity density than is observed towards the fainter end
\citep{Bethermin2022, Bisigello2022}.  While it appears there is a
tantalising break in the CO(4--3) LF we find it is not statistically
significant (see \autoref{sec:lumin-funct-analys}) and the differences
could again be caused by cosmic variance.  If real, such a break could
be caused by a rapid change in the average excitation between the
faint and bright end of the LF at these redshifts, though such a
scenario is not seen in the simulations, which are otherwise
consistent with the LF in the HDF-N.

As the cosmological deep fields are biased \emph{against} having very
low-redshift galaxies in the foreground, we only provide a $3\sigma$
upper limit on the CO(1--0) luminosity function at the bright end of
$\log \phi\,[\mathrm{Mpc}^{-3}\,\mathrm{dex}^{-1}]\leq -1.9$.  This is
somewhat more stringent than that of ASPECS due to the increased
volume.  There are no high-fidelity sources contributing to CO(6--5)
at the highest redshifts, implying a $3\sigma$ upper limit at the
bright end of the luminosity function of
$\log \phi\,[\mathrm{Mpc}^{-3}\,\mathrm{dex}^{-1}]\leq -3.6$.  We do
note that the \textsc{eazy} photometric redshifts do not fully cover the
redshift range spanned by CO(6--5), which implies some signal may be
lost in this bin (though the majority of the signal is expected to
come from the no-counterpart sources), hence the constraints should be
viewed as conservative.  The same upper limit also extends to the
higher-$J$ lines, nearly independent of transition (cf.\
\autoref{fig:colf}).  For both CO(1--0) and CO (6--5), the upper
limits are comfortably in agreement with theoretical models.

The original PdBI pointing \citep{Walter2014, Decarli2014} was chosen
specifically to include the bright sub-millimeter galaxy HDF\,850.1
and hence could not provide constraints on the CO(5--4) LF.  The full
mosaic, however, is not chosen to include this source.  Therefore, it
now provides the first measurement of the CO(5--4) LF at
$\avg{z} = 5$.  The constraints on the LF are still limited and
subject to cosmic variance (given that the expected source density of
galaxies like HDF\,850.1 is $\lesssim 1$ for the area of the HDF-N
survey, e.g., \citealt{Zavala2021}), but in overall agreement with the
predictions from \textsc{Spritz}.

In contrast to the empirical models discussed above,
\cite{Popping2019} have used the IllustrisTNG hydrodynamical
simulations \citep{Weinberger2017, Pillepich2018b} and the Santa Cruz
semi-analytical model \citep{Somerville1999, Somerville2001} to
demonstrate that simulations which model galaxies from first
principles generally do not reproduce the observed luminosity
functions, unless the HUDF is a strong outlier (i.e., only a small
fraction of the various realisations agreed with the HUDF).  The
result that our best-constrained CO LFs in the HDF-N, CO(2--1) and
CO(3--2), do not deviate strongly from those the HUDF implies there is
increasing tension with the simulations that predict significantly
smaller amounts of molecular gas in galaxies on average
\citep{Popping2019}.  These simulated CO LFs are quite sensitive to
the assumed prescription for the value of \aco, that is needed to
convert the simulated molecular gas mass function to a CO LF.  While
better agreement can be found by assuming significantly lower average
values of $\aco \lesssim 1$, tension then still remains in matching
the observed faint- and bright end of the mass and luminosity
functions simultaneously \citep[e.g.,][]{Popping2019, Dave2020}.

\subsection{Cosmic molecular gas density and cosmic variance}
\label{sec:cosmic-molecular-gas}
\begin{figure}[t]
\includegraphics[width=\columnwidth]{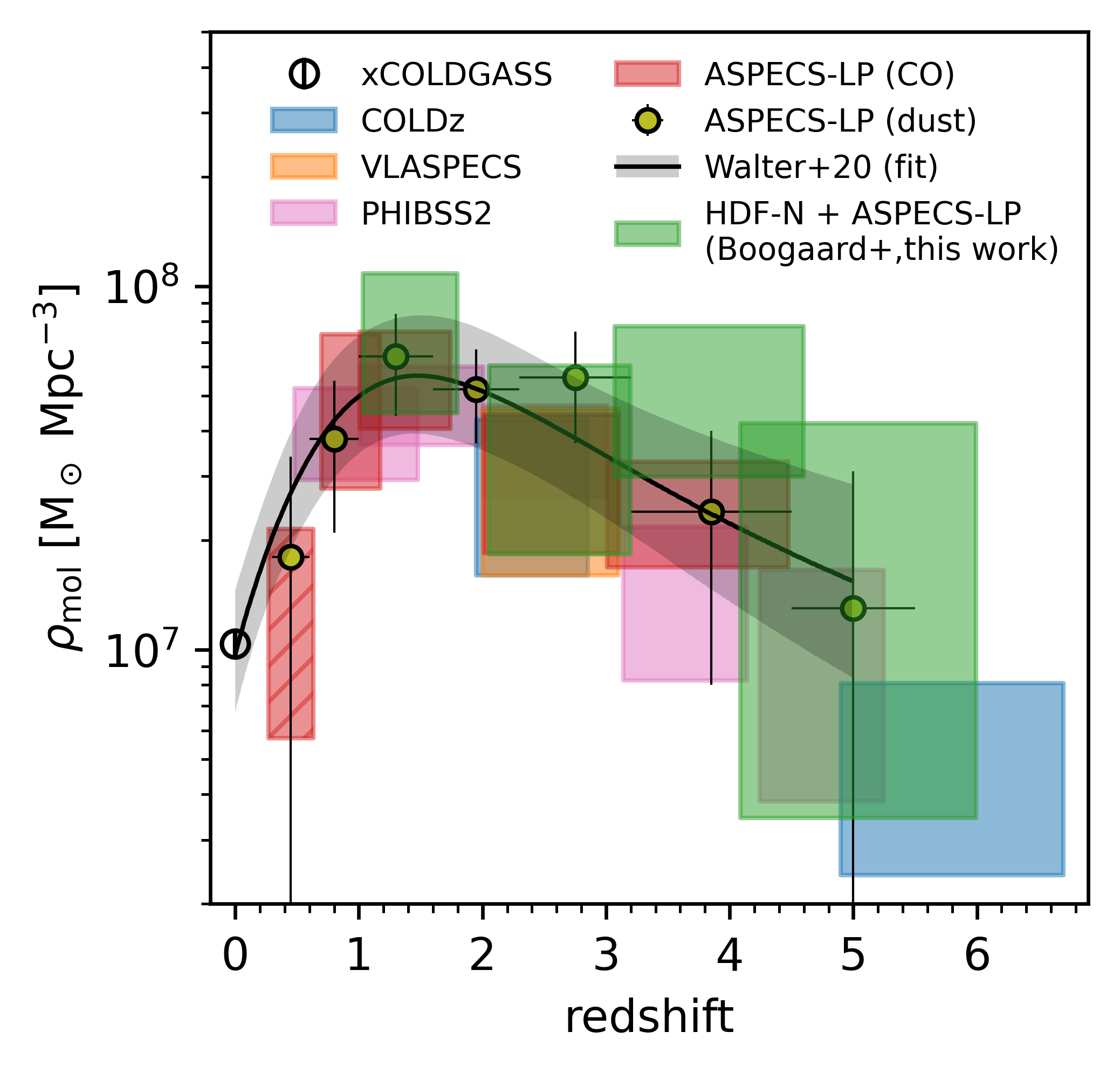}
\caption{Cosmic molecular gas density from integrating the CO
  luminosity function fits for the HDF-N + HUDF (green boxes).  We
  compare this to the local measurements from xCOLDGASS
  \citep{Fletcher2020}, COLDz in COSMOS and GOODS-N \citep{Pavesi2018,
    Riechers2019}, VLASPECS in the HUDF \citep{Riechers2020b}, PHIBSS2
  background sources \citep{Lenkic2020}, and the measurements from
  ASPECS in the HUDF \citep{Decarli2019, Decarli2020}, incl.\ the dust
  continuum stacking \citep{Magnelli2020}.  The gray line shows the
  join fit to all the literature data from \cite{Walter2020}.  The
  joint measurements confirm a rise of the cosmic molecular gas
  density from the local universe by a factor 4.5--11$\times$ to
  $z\sim1.5$ and subsequent decline out to higher redshift, with
  slightly larger uncertainties reflecting the field-to-field variance
  between the HUDF and HDF-N.  \label{fig:rhoh2}}
\end{figure}
The evolution of the cosmic molecular gas density is determined from
the CO LFs at different redshifts.  We combine the constraints from
NOEMA in the HDF-N and ASPECS in the HUDF, by integrating the
Schechter fits to both surveys down to the lowest $L'$ probed in the
data (i.e., we do not extrapolate the faint end).  This is consistent
with earlier work and the integral should trace the bulk of the
molecular gas mass given that the knee of the luminosity function is
well sampled.  Indeed, the stacking results at $1\leq z < 2$ from
ASPECS \citep{Inami2020} imply that there is not a large amount of gas
mass missed at fainter luminosities.  To convert the observed CO
emission to a gas mass one needs to adopt a CO-to-molecular gas
conversion factor, \aco\ (including He), and average excitation
correction that is representative for galaxies around the knee of the
LF.  We assume an
$\aco = 3.6$\,\Msun(K\,km\,s$^{-1}$\,pc$^{2}$)$^{-1}$
\citep{Daddi2010, Bolatto2013}, mainly to be consistent with earlier
work, but note all results can simply be linearly rescaled to a
different \aco.  For the excitation, we adopt the average values
derived for galaxies at $z<2$ and $z>2$ from \cite{Boogaard2020}, with
$r_{J1} = [0.75, 0.80, 0.61, 0.44]$ for $J = [2, 3, 4, 5]$.  These ratios
have been derived from the flux-limited sample of CO-emitters from the
ASPECS survey and were also used for the ASPECS measurements
\citep{Decarli2020}.  Note the assumed $r_{51}$ is also reasonably
similar to the excitation measured in HDF\,850.1, with
$r_{52} \approx 0.47 \pm 0.15$ \citep[][assuming that CO(2--1) will be
highly excited in the intense starburst, even if only from the high
CMB temperature at this redshift]{Walter2012}.  The resulting cosmic
molecular gas densities are shown in \autoref{fig:rhoh2} and tabulated
in \autoref{tab:rhomolparams}.

\begin{deluxetable}{lcc}[t]
  \tablecaption{Cosmic molecular gas density from the NOEMA HDF-N + ASPECS HUDF ($1\sigma$ range).
    \label{tab:rhomolparams}}
  \tablehead{
\colhead{Transition}  & \colhead{$\langle z \rangle$} & \colhead{$\log \rho_{\rm mol}$}   \\
\colhead{}            & \colhead{}                    & \colhead{(M$_{\odot}$\,Mpc$^{-3})$} \\
\colhead{(1)} & \colhead{(2)} &\colhead{(3)}
} \startdata
CO(2--1)              & 1.4389                        &  7.65, 8.04    \\
CO(3--2)              & 2.6276                        &  7.26, 7.78    \\
CO(4--3)              & 3.8210                        &  7.48, 7.89    \\
CO(5--4)              & 5.0156                        &  6.54, 7.62    \\
\enddata
\tablecomments{The molecular gas densities, $\rhomol$, are given as the 16th and 84th percentiles ($1\sigma$).}
\end{deluxetable}

The new constraints on the cosmic molecular gas density at $1<z<1.8$
and $2<z<3.1$ are in good agreement with earlier measurements,
including those derived (independently) from the dust continuum in the
HUDF \citep{Magnelli2020}, as well as the constraints from lower-$J$
transitions at the same redshift in COSMOS, GOODS-N and the HUDF
\citep{Pavesi2018, Riechers2019, Riechers2020b}.  The uncertainties
are similar or in fact somewhat larger than the measurement from
ASPECS alone at $\avg{z} = 1.4$, which reflects the impact of cosmic
variance on the LF and subsequent \rhomol\ measurements.  The combined
constraints confirm the rise in the cosmic molecular gas density from
redshift zero out to $z \sim 1.5$, with a factor between
4.5--11$\times$, and a subsequent decline out to higher redshift.

The relatively high $\rhomol(3<z<4.5)$ is because the best-fit
Schechter function appears to somewhat overestimate the knee of the
LF.  As stated in \autoref{sec:co-lumin-funct}, the constraints on the
CO(4--3) LF are rather limited, and we conclude that there is still
significant uncertainty in the detailed shape of the LF as well as the
molecular gas density at these redshifts.  Indeed, the measurements
from \citep{Lenkic2020} fall significantly below the average
\citep{Walter2020}.  Alternatively, it could be that the average
excitation is higher than assumed, though this is unlikely to explain
the full discrepancy.  Further analysis is needed to better constrain
the luminosity density at these redshifts \citep[cf.][]{Boogaard2021}.
We also show the constraints on $\rhomol(4<z<6)$ based on the CO(5--4)
LF.  The large uncertainties are due to the limited constraints on the
LF, though overall the value are consistent with early measurements.
The scatter around the average excitation is expected to be
significantly larger for the higher-$J$ lines that trace warmer and
denser gas, which makes the estimate of the associated total gas mass
from these lines more uncertain.

The new measurements from the NOEMA HDF-N survey provide an important
complement to the earlier measurements in other fields.  For both the
luminosity function and molecular gas density, perfect agreement is
not expected because these are determined from different fields.  The
overall good agreement between the LFs in the HDFN and HUDF,
especially for CO(2--1) and CO(3--2) implies that the impact of cosmic
variance on these LFs is not more severe than previously estimated
\citep{Popping2019, Decarli2020}.  Indeed, while the area-on-sky of
the spectral scan surveys are typically modest, the broad redshift
coverage implies that substantial volumes are probed (cf.\
\autoref{tab:freqzvol}), mitigating the impact of field-to-field
variance \citep{Popping2019}.  The combined measurements from the
HDF-N and HUDF presented here now fold in the uncertainties due to
cosmic variance between the two fields directly.

\section{Conclusions} \label{sec:conclusion}

This paper presented an 8.5 arcmin$^{2}$ NOEMA survey of the Hubble
Deep Field North (HDF-N), that scans nearly the complete 3\,mm band
(from 82--113\,GHz) band in 45\,pointings, to identify for molecular
line emission in distant galaxies, measure the CO luminosity functions
(LF), and constrain the cosmic molecular gas density (\rhomol) out to
$z\sim6$.  The main conclusions of this study are as follows.

\begin{enumerate}

\item We search for line candidates in the cube via matched filtering
  and determine a redshift probability distribution, $p(z)$, for each
  of the line candidates exploiting the existing photometric redshift
  distributions of nearby counterparts in combination with a CO
  redshift prior, including a no-counterpart solution.  Out of the
  larger sample of candidates, we identify 7 high-confidence line
  emitters (with $\SN \geq 5.85$ and fidelity $F>0.7$).  Four are
  CO(2--1) emitters at $1<z<2$ (of which three spectroscopically
  confirmed), two have a broader $p(z)$ but are most likely CO(3--2)
  emitters at $2<z<3$, and the final source is HDF\,850.1, a
  well-known starburst galaxy at $z=5.184$ \citep{Walter2012} detected
  in both CO(5--4) and CO(6--5).

\item We detect four high-confidence 3\,mm continuum sources.  One is
  HDF\,850.1, while the other three are all identified as known radio
  galaxies with spectroscopic redshifts $z\leq1$.

\item The larger area and significant depth of the NOEMA HDF-N survey
  compared to earlier studies provides the first constraints on the
  bright end of the CO LFs for $J=2$ up to $5$ at $1 < z < 6$,
  extending the existing LF measurements up to $z\sim4$ from the knee
  upwards.  We find a lower density in the overlap region near the
  knee of the CO(2--1) and CO(3--2) LFs in the HDF-N compared to the
  HUDF (from ASPECS) of $\sim 0.15$ and 0.4 dex, respectively.  We
  find tentative evidence for a higher CO(4--3) luminosity density at
  the bright end than expected from extrapolations of earlier surveys,
  though in good agreement with simulations.  Finally, we provide the
  first constraints on the CO(5--4) LF at $\avg{z} = 5$.

\item We perform a joint analysis of the LFs in the HDF-N and HUDF
  (from ASPECS) and find that they are well described by Schechter
  functions up to at least $J=3$.  Given that the constraints were
  determined from two completely independent fields, this suggests
  that the current measurements of the LFs and subsequently the cosmic
  molecular gas density are not strongly affected by cosmic variance.

\item The agreement between the HDF-N and HUDF poses some challenges
  for simulations that model galaxies from first principles, that
  (under the assumption of an \aco) generally predict lower values for
  the CO LFs than are observed.

\item We integrate the combined HDF-N and HUDF LFs to provide revised
  constraints on the molecular gas density from the joint fields.  The
  uncertainties on \rhomol\ from the joint determination are similar
  and in some cases even slightly larger.  The latter is a direct
  consequence of the field-to-field variance which is now reflected in
  the measurements.  We find very good agreement with earlier surveys
  for $\rhomol(1.0 < z < 1.8)$ and $\rhomol(2<z<3.2)$ and
  $\rhomol(4<z<6)$.  The results show that the cosmic molecular gas
  density increases by a factor 4.5--11 from redshift 0 to
  $z\sim1.5$, in agreement with previous measurements including
  independent measurements from the dust continuum.

\end{enumerate}

The independent constraints from the NOEMA HDF-N survey provide
important constraints on the cosmic variance in the CO LF compared to
the other deep fields such as the HUDF.  On-going efforts such as WIDE
ASPECS are expanding the spectral scan surveys to even larger areas to
further constrain the variance in the bright end of the CO LF (not
covered by ASPECS).  The key combination of depth and relatively large
area of the NOEMA HDF-N survey is made possible by the increased
sensitivity of the extra antennas and in particular the large
instantaneous bandwidth, allowing to scan the entire 3\,mm band in
only 2 tunings.  Future upgrades on NOEMA, such as the dual-band,
full-band and multi-beam receivers, as well as the upcoming Band\,2
and bandwidth upgrades for ALMA \citep{Carpenter2022}, would allow to
constrain the evolution of the cosmic molecular gas density even
further.

\begin{acknowledgments}
  We want to thank the referee for a constructive and helpful report.
  We are grateful to Matthieu B\'{e}thermin and Laura Bisigello for
  providing the simulated CO luminosity functions from SIDES and
  \textsc{Spritz}, respectively.  M.A. acknowledges support from
  FONDECYT grant 1211951, CONICYT + PCI + INSTITUTO MAX PLANCK DE
  ASTRONOMIA MPG190030, CONICYT + PCI + REDES 190194, and ANID BASAL
  project FB210003.  R.S.E. acknowledges funding from the European
  Research Council (ERC) under the European Union’s Horizon 2020
  research and innovation programme (grant agreement No 669253).  This
  work is based on observations carried out under project numbers
  W18DI and W19CR with the IRAM NOEMA Interferometer. IRAM is
  supported by INSU/CNRS (France), MPG (Germany) and IGN (Spain).
\end{acknowledgments}

\vspace{5mm}
\facilities{NOEMA (IRAM)}

\software{\textsc{Topcat} \citep{Taylor2005}, \textsc{Gnuastro}
  \citep{Akhlaghi2015}, \textsc{IPython} \citep{Perez2007},
  \textsc{numpy} \citep{Harris2020}, \textsc{scipy}
  \citep{Virtanen2020}, \textsc{Matplotlib} \citep{Hunter2007},
  \textsc{Astropy}, \citep{TheAstropyCollaboration2018},
  \textsc{Interferopy} \citep{interferopy}, \textsc{qubefit}
  \citep{qubefit}, \textsc{Ultranest} \citep{Buchner2021}.}

\appendix
\newpage
\section{Luminosity functions: impact of methodology, priors and
  completeness}
\label{sec:colf-impact}
\begin{figure*}[t]
  \centering
  \includegraphics[width=0.9\textwidth]{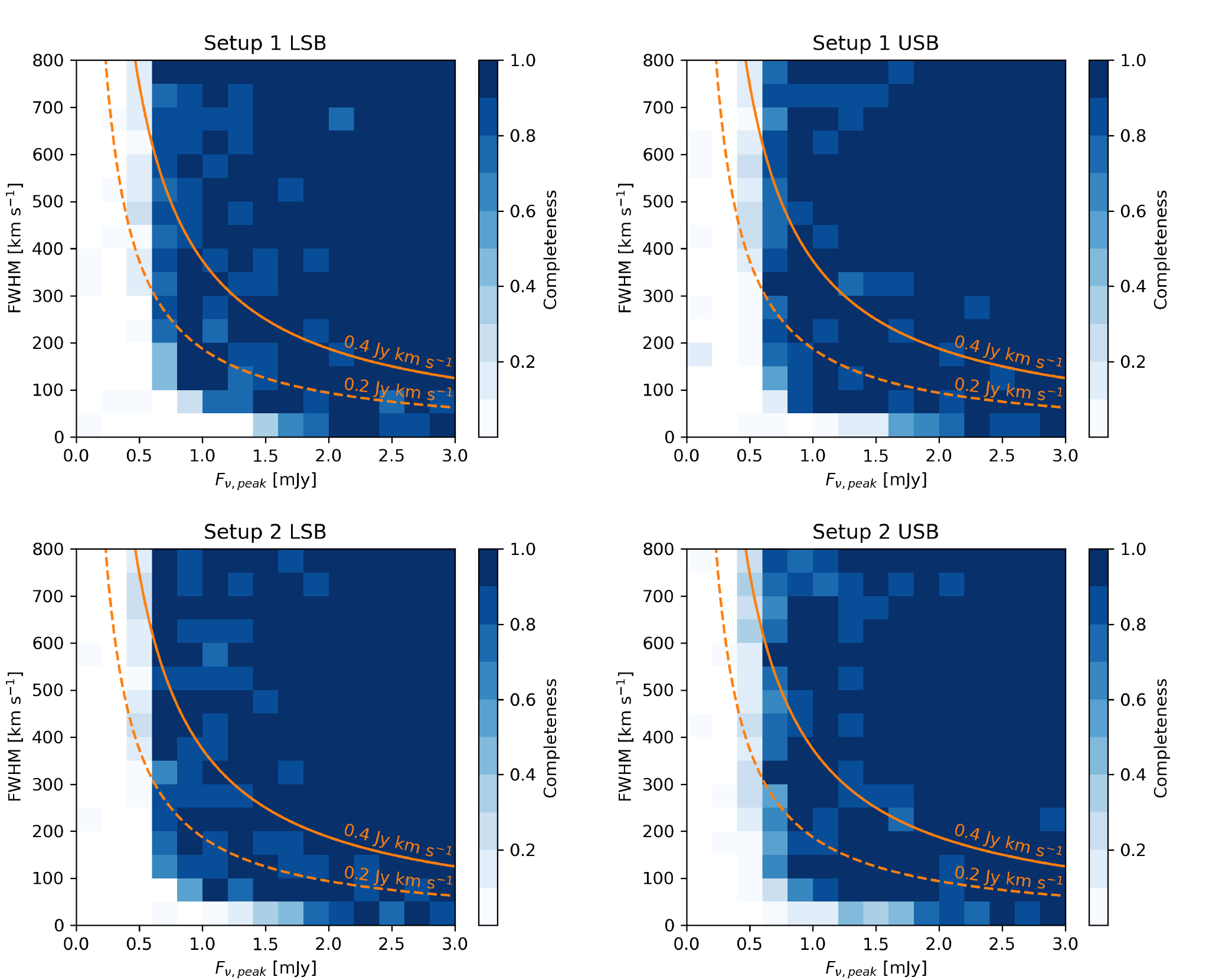}
  \caption{Completeness fractions for the four fields determined from
    injecting and recovering mock sources for a range of peak fluxes
    and line widths.  The completeness is similar between the four
    sidebands, with a slightly higher completeness for the sidebands
    with lower rms.  Overall, the completeness exceeds $90\%$ for
    integrated line fluxes above roughly 0.4\,Jy\,km\,s$^{-1}$, being
    higher for narrower lines at fixed integrated
    flux. \label{fig:compl}}
\end{figure*}
\begin{figure*}[t]
\includegraphics[width=\textwidth]{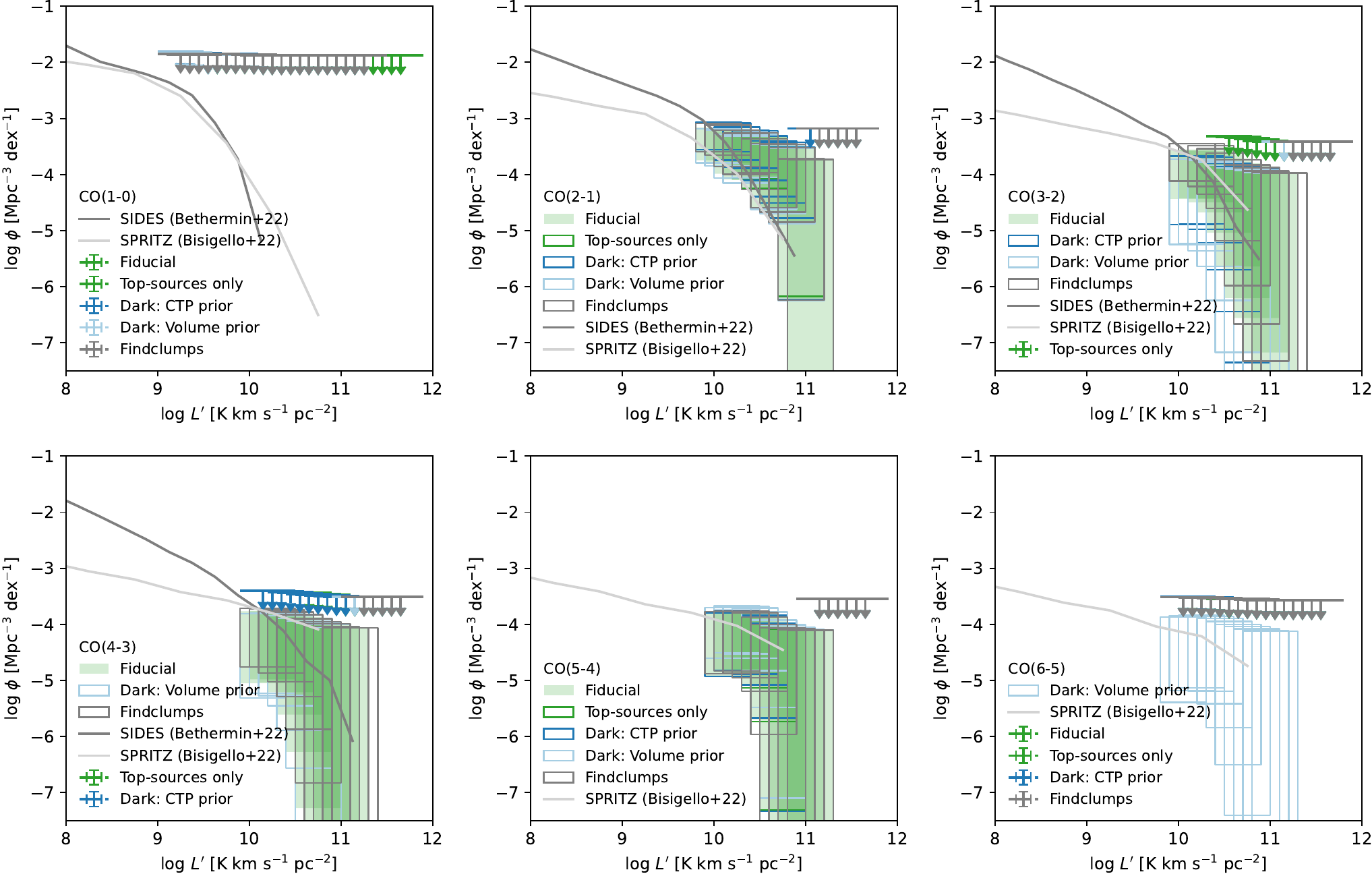}
\caption{Impact on the CO LFs of different codes and assumptions,
  compared to the fiducial result from the paper (\autoref{fig:colf}).
  The result obtained when using the line candidates from
  \textsc{Findclumps} (grey) are nearly identical to those obtained
  with MF3D, which is not only reassuring, but also supports the join
  analysis with ASPECS.  Including only the top sources ($F>0.7$; dark
  green) leaves the $J=2$ and 5 LFs unchanged, but turns the
  constraints for $J=3$ and 4 into upper-limits only. The effect of
  changing the CO-line prior for the sources without counterpart to
  that of the sources with counterpart (dark blue) lowers the
  constraints on $J=4$, with minimal impact on the other lines.  Using
  the volume prior used in \cite[][light blue]{Decarli2019} instead,
  boosts the $J\geq5$ LFs in favor of the lower-$J$
  lines.  \label{fig:colf-alt}}
\end{figure*}
\begin{deluxetable}{@{\extracolsep{4pt}}ccccc@{}}[t]
  \tablecaption{CO luminosity functions. \label{tab:colfs}}
  \tablehead{
    & \colhead{CO(2--1) at $\avg{z} = 1.4389$} & \colhead{CO(3--2) at $\avg{z} = 2.6276$} & \colhead{CO(4--3) at $\avg{z} = 3.821$} & \colhead{CO(5--4) at $\avg{z} = 5.0156$} \\
    \cline{2-2} \cline{3-3} \cline{4-4} \cline{5-5}
\colhead{$\log L'_{*}$} &  \colhead{$\log \phi_{*}$} &  \colhead{$\log \phi_{*}$} & \colhead{$\log \phi_{*}$} & \colhead{$\log \phi_{*}$} \\
\colhead{(1)} & \colhead{(2)} &\colhead{(3)} & \colhead{(4)} &\colhead{(5)}
}
\startdata
10.05 & -3.75, -3.16 & \nodata       & \nodata       & \nodata  \\
10.15 & -3.79, -3.18 & -4.44, -3.56  & -5.05, -3.77  & -4.94, -3.79 \\
10.25 & -3.99, -3.25 & -4.43, -3.56  & -4.98, -3.76  & -4.84, -3.76 \\
10.35 & -4.09, -3.29 & -4.50, -3.59  & -5.03, -3.77  & -4.85, -3.76 \\
10.45 & -4.01, -3.27 & -4.68, -3.64  & -5.21, -3.81  & -4.90, -3.78  \\
10.55 & -4.22, -3.35 & -5.04, -3.72  & -5.61, -3.87  & -5.09, -3.84 \\
10.65 & -4.57, -3.44 & -5.63, -3.80  & -6.26, -3.93  & -5.70, -3.99  \\
10.75 & -4.67, -3.46 & -6.20, -3.84  & -7.24, -3.96  & -7.40, -4.06  \\
10.85 & -4.85, -3.52 & -6.54, -3.88  & -7.86, -4.00  & -15.81, -4.08\\
10.95 & -6.26, -3.72 & -7.16, -3.94  & -8.48, -4.04  & -17.02, -4.10 \\
11.05 & $\leq$ -3.18 & -8.63, -3.97  & -9.78, -4.06  & -18.77, -4.10 \\
11.15 & \nodata      & $\leq$ -3.42  & $\leq$ -3.51  & $\leq$ -3.55  \\
\enddata
\tablecomments{(1) Center of the 0.5\,dex-wide log luminosity bins.
  (2-5): Luminosity functions; the values denote the 16th and 84th
  percentile ($1\sigma$) or a $3\sigma$ upper limit.}
\end{deluxetable}

In this section we investigate the impact of the methodology
assumptions made for the CO-line identification and the subsequent
impact on the luminosity function and cosmic molecular gas density.
The completeness corrections as a function of line width and peak flux
for each of the four cubes (discussed in \autoref{sec:linesearch}) are
shown in \autoref{fig:compl}.  Overall, we find the completeness
corrections are minor, as the uncertainties are dominated by the
sample purity (fidelity) and not completeness.

While the different line-search code typically agree on the
high-fidelity candidates, there are increasing differences in the
exact number of candidates and their \SN\ and Fidelity for fainter
lines.  We therefore repeat the full CO LF analysis using the line
candidates from \textsc{Findclumps} instead of MF3D.  The results are
shown in \autoref{fig:colf-alt}.  Reassuringly, the CO LFs are
effectively unchanged, with the only difference being a slightly
higher count for the faintest bins.  To check whether our computation
of the LF is consistent with earlier work, we also recompute the
CO(2--1) and CO(3--2) LFs from the ASPECS large program using the top
15 high-fidelity line candidates \citep{Gonzalez-Lopez2019} and find
that we recover the LFs from \cite{Decarli2019} perfectly (up to the
completeness corrections at the faint end).  We also investigate the
impact of the prior on the sources with out a photometric counterpart.
The first alternative we explore is to assume the same prior as for
sources that do have a counterpart.  This redistributes the lines
towards the CO(2--1) bin, slightly reducing in the CO(3--2) bin, and
loses constraints on the CO(4--3) bin (leaving only upper limits),
whilst leaving CO(5--4) unchanged.  We also explore take sources
without counterpart into account with a weight purely proportional to
the volume probed at each redshift (as in \citealt{Decarli2019}). This
has the opposite effect, of reducing the $J\leq4$ transitions more
strongly towards lower-$J$, whilst slightly boosting the $J\geq5$
luminosity functions, such that there are sufficient counts also in
the $J=6$ LF (i.e., effectively more than one source).

In summary, we find the CO(2--1) and CO(5--4) luminosity functions are
robust, because they are well-defined by the sources that have a
confident identification, such that the prior assumptions (or
different line search codes) have minimal impact on the final result
to within uncertainties.  Moreover, the impact on the (combined)
$\rhomol$ constraints for these lines are negligible.  For CO(3--2)
and CO(4--3) somewhat larger variation is seen towards the faint end,
while the bright end remains robust.  Using only the top-fidelity
sources, however, results in only upper limits on CO LFs for the
latter two transitions.
\section{Redshift distributions}
\label{sec:redsh-distr}

\begin{figure*}[t]
  \vspace{-1cm}
  \includegraphics[width=\textwidth]{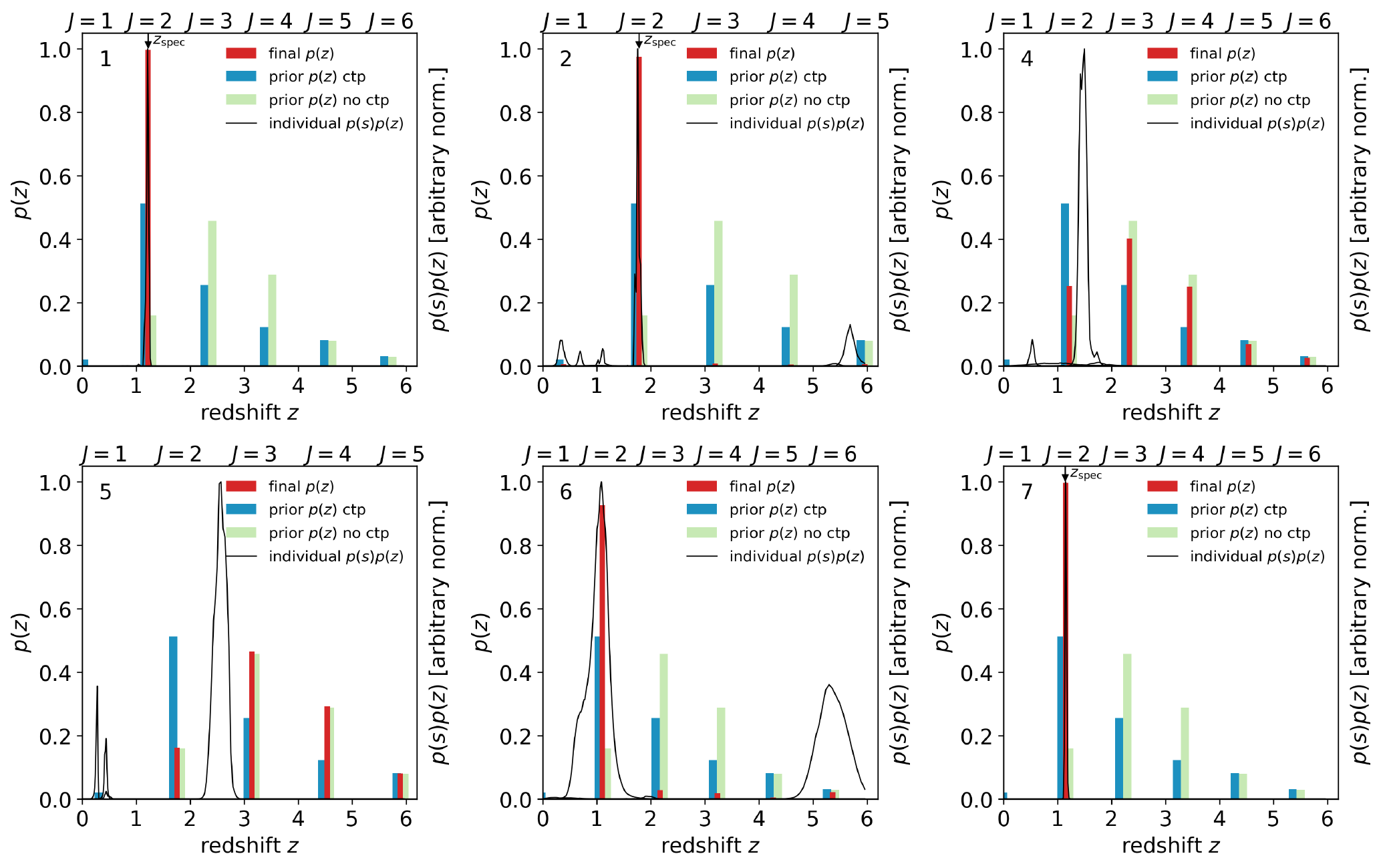}
  \caption{Redshift probability distributions, $p(z)$, for the top
    candidates (except HDF\,850.1).  The solid red bars show the final
    $p(z)$ and the other bars show the priors
    (cf.\ \autoref{fig:priors}).  The black lines show the photometric
    redshift distributions of the individual sources weighted by the
    separation, $p(s)p(z)$ (with arbitrary normalisation between the
    panels). The black arrow in the top of each panel shows the
    spectroscopic redshift for the most-likely photometric counterpart
    (if available)---in all cases coincident with the peak of the
    $p(z)$. \label{fig:pz}}
\end{figure*}

The redshift probability distributions (\autoref{eq:pz}) for the top candidates
from \autoref{tab:linesearch} are shown in \autoref{fig:pz}, except
for HDF\,850.1 which is known to have no photometric counterpart and a
spectroscopic redshift of $z=5.184$ \citep{Walter2012}.
\clearpage
\bibliography{library}{}
\bibliographystyle{aasjournal}

\end{document}